# On the pole structures of the disconnected part of hyper elliptic g loop M point super string amplitudes

A.G. Tsuchiya.

Oct 7, 2012   Revised Apr 3, 2015

Abstract

Structures of the disconnected part of higher genus superstring amplitudes restricted to the hyper elliptic cases are investigated in the NSR formalism, based on the D'Hoker Phong and recent results [1]-[5]. A set of equations, which we can regard as a basic tool to sum over the spin structures of any of g-loop, M point amplitudes systematically, is shown by using a classical result of Abelian functions. We discuss structures of g-loop, M-point massless external boson superstring amplitudes by assuming that the spin structure dependence of any integrand of the disconnected amplitudes, excluding measure part, is only on one kind of constants: the genus g Weierstrass Pe function valued at the summation of g number of half periods chosen out of 2g+1 half periods. This is a natural generalization of the case of genus 1. This assumption will be validated by a conjectured theorem which states that the spin structure dependent part of any string amplitude integrand will be naturally decomposed into two parts: one is composed of manifestly modular invariant functions of positions of inserting operators, and the other is the polynomial of Pe function constants related to the moduli of Riemann surfaces only. It is shown that this is actually the case for any M for g=1, and M=1,2,3 for any g. Due to a technical problem, our consideration is at present restricted to the case that g(g+1)/2 is odd. Example calculations are shown for the genus 2 by the method described here. In particular, our method correctly reproduces biholomorphic 1 form of D'Hoker Phong result[1] as for the four point amplitudes of the disconnected parts.

## 1. Introduction

The investigations of the structures of the string amplitudes have a long history by various methods. After the breakthrough papers by D'Hoker Phong on the two loop measures and amplitudes , construction of the superstring measure for arbitrary genus g has been one of the important issues[1]-[22]. In the hyper elliptic case, the substantial difficulty on the super string measure is absent, and the measure is given for any g in the paper of [5]. Rather surprisingly, as shown by DHP, in the massless 4 point amplitude of genus 2, the disconnected part of the amplitude is found to be cancelled with a part from the connected part. As is always the case, in the NSR formalism, we face quite complicated calculations on summing over the spin structures, apart from the substantial problems on the moduli.of string measures.

If, as a preliminary consideration, we restrict ourselves to consider the calculations of hyper elliptic, disconnected parts of the amplitudes, it is possible to construct a rather general method of summing over spin structures, and therefore construct methods of calculating higher points amplitudes in general genus, with the help of a classical result on Abelian functions. This is our main result, shown in the equations (4.50), (4.51).

Our considerations are based on one natural assumption. It is that any integrand of the disconnected M point massless amplitudes in any genus g, excluding string measure, depends on the spin structures only through one kind of constants, the genus g Weierstrass Pe function valued at the summation of g number of half periods chosen out of 2g+1 half periods. Each choice of g number of half periods corresponds to one spin structure. This kind of constants is denoted as $P_{IJ}(\frac{1}{2}\Omega_\delta)$ in the text. Because at present we have rigorous proof of this validity only for any M in g=1 and for M=1,2,3 in g>1, our calculation method which will be described in the following sections is only reproducing the results we already know by a different method. However our method described in section 3 and 4 is systematical and it will be a good tool when we calculate higher point amplitudes in higher genus and it can be applied to any genus g and M point amplitudes. We will see as an example the possibility of calculating the g loop four point amplitude in the disconnected part and discuss what can be said about that amplitude in section 6.

The author expects that this assumption will be validated in a much stronger form by a conjectured theorem for the genus g. It states that the spin structure dependent part of any disconnected string amplitude will be decomposed into the summation of the products of two kinds of factors: one is composed of manifestly modular invariant functions of positions of inserting operators, and the other is the polynomial of Pe function constants related to the moduli of Riemann surfaces only. The former will be differentials of the unique function : genus g sigma function. This is a generalization of (1.4) or (2.18), which is proved for g=1 sometime ago.   We will review it in the section2.

If we try to seriously calculate general g loop, M point amplitude for the hyper elliptic case, and if succeeded, what will be the contents in its final form after summing over all spin structures?   Obvious contents are Green functions resulted from the contractions of bosonic operators in the vertexes , contracted products of momentum and polarization vectors, and products of holomorphic 1 forms defined on the Riemann surface, zero mode contributions in case of closed strings.   These are combined with a result of contractions of fermion   fields in the vertices obtained after summing over all spin structures using appropriate string measures.   The result of contractions of fermion fields in the vertex operators will include two parts: one is a combination of modular invariant functions of vertex inserting points expressed by *unique* theta function which should show clear pole structures of the amplitudes and should have good properties under the modular transformation.   And the other will be constant modular forms which are expressed as symmetric functions of the branch points of the curve, $e_1, e_2, ..... e_{2g+1}$.

The general calculation of all of actual contractions including the elements described above is not the issue pursued in this document because it is too cumbersome and it contains a lot of calculations which are not substantial.    Instead, we would like to see what kind of patterns appear in the calculation of the amplitudes on some important aspects, especially on the fermion field contractions.   After we clarify or simplify the methods of taking over the spin structures, much will be understood about the general pole structure of the string amplitudes as we will see.

A contraction of 2M number of fermion fields:

$$< \prod_{i=1}^{M} \varsigma_i \cdot \psi(z_i) k_i \cdot \psi(z_i) > \qquad (1.1)$$

leads to the M number of simple products of Szego kernels of the form in the disconnected part of the amplitudes:

$$S_\delta(z_1, z_2) S_\delta(z_2, z_3) ..... S_\delta(z_M, z_{M+1}) \qquad \text{with} \quad z_{M+1} = z_1 \qquad (1.2)$$

where $z_1, z_2 .... z_M$ are the inserting points of vertex operators on the Riemann surface, $\delta$ represents a spin structure.   For the higher genus g, appropriate Abel maps are assumed.

Along our original motivations described above, the structure of this form will be one of our interests in this document. In the standard conformal field theory, where fermion field contractions give a determinant of Szego kernels, it is often said that the Fay's formula relates fermion contractions to boson contractions. On the other hand, in calculating superstring amplitudes, this picture does not work so naively because each of the fermion field contractions is multiplied with complicated momentum and polarization vectors of the external bosons. The point is that, if we expect there must be nice theta function identities which make it possible to sum over spin structures, we have to consider the spin structure on *each* of the product of the form (1.2). This is the reason why we consider the *simple* product of the Szego kernels. The key is the cyclic condition $z_{M+1} = z_1$ as we will see.

The elliptic curve we use is of the form

$$s^2 = \prod_{k=1}^{2g+1}(x - e_k) = R(x) \tag{1.3}$$

where one of the branch points $e_{2g+2}$ is fixed at $\infty$.

In many literatures, the following form of the Szego kernels is used : A spin structure corresponds to the grouping the 2g+2 branch points $e_1, e_2, \ldots e_{2g+2}$ into two sets, $\{a_1, a_2, \ldots a_{g+1}\}$ and $\{b_1, b_2, \ldots b_{g+1}\}$. The Szego kernel is given by

$$S_\delta(x, y) = \frac{1}{2} \frac{s_A(x) s_B(y) + s_A(y) s_B(x)}{x - y} \left(\frac{dx}{s(x)}\right)^{\frac{1}{2}} \left(\frac{dy}{s(y)}\right)^{\frac{1}{2}}$$

where

$$s_A(x)^2 = \prod_{i=1}^{g+1}(x - a_i) \qquad \text{and} \qquad s_B(x)^2 = \prod_{i=1}^{g+1}(x - b_i)$$

Actually, this often gives a very convenient way of calculation and lead to important results. But here we do not adopt this form of Szego kernel, since we want to keep modular invariant theta function expressions on the operator inserting points in the final form of the amplitudes, aiming to clarify the structure of higher point calculations in higher genus.

Instead, we will pursue the way something like the following, in analogy of the results in g=1. In g=1, where all amplitude calculations correspond to the disconnected parts, there is a kind of "decomposition" formula of the product of Szego kernels [23]. It is schematically written as

$$\prod_{i=1}^{M} S_\delta(z_i - z_{i+1}) = \sum [elliptic\ functions\ of\ z_1,....z_M, V] \cdot [\exp onent\ of\ the\ branch\ po\text{int}\ s\ e_\delta]$$

(1.4)

for the arbitrary number of M. Each of V, which will be defined in section 2, is a manifestly elliptic function which shows explicit pole dependence of the amplitudes as a function of vertex inserting points $z_1, z_2....z_M$. The V is totally written by the unique odd theta function on torus, theta-1 in the classical notation, and it remains unchanged in the whole process of the spin structure summation. In higher point amplitudes its explicit form will be included in the final expression unless the whole vanishes by the spin structure summation on the other part.

The spin structure dependence of the amplitude is only in the exponent of the branch points, $e_\delta$, which does not depend on operator inserting points $z_1, z_2....z_M$.

Once this formula is applied, the summation over spin structures can be done only algebraically on the moduli dependent parts, and in general it gives symmetric function of the branch points. This formula contains infinitely many theta identities which are precisely sufficient to sum over spin structures for arbitrary M as reviewed in section 3. The final form of the amplitude can be written in a manifestly modular invariant way, which are all included in V from the beginning, and the pole structure of the amplitude is completely clarified for any value of M in the sense explained above.

This kind of decomposition is also seen for M=1,2,3 in genus g as we will see later. It is expected that this formula can be generalized to the higher genus surfaces for general M, but this is a future issue.

At g=1, as is well known, the three branch points $e_\delta$ is related to the value of the Weierstrass' Pe function at the half periods, denoted as $\omega_\delta$: $e_\delta = P(\omega_\delta)$. In string theories, the $\omega_\delta$ are often denoted as $\frac{1}{2}, \frac{\tau}{2}, -\frac{1+\tau}{2}$. In section4, the reason why we expect the spin structure dependence of the amplitudes except the measure part is only on the constants $P_{IJ}(\frac{1}{2}\Omega_\delta)$ in genus g, which are generalizations of $P(\omega_\delta)$, is explained. The indices I and J are from 1 to g, coming from the definition of Pe function at higher genus g. For g>1, a non-trivial problem occurs which did not exist in g=1. At a fixed choice of spin structure $\delta$, there are $\frac{g(g+1)}{2}$ number of constants in $P_{IJ}(\frac{1}{2}\Omega_\delta)$ which are symmetric on $I, J (=1,2,....g)$. The $\delta$ varies $\delta = 1,2,......,\binom{2g+1}{g}$, until the number equal to that of non-singular even characteristics among $2^{g-1}(2^g+1)$ even characteristics. In g=1, I=J=1 and $\delta = 1,2,3$ and the value of all $P_{IJ}(\frac{1}{2}\Omega_\delta)$ are easily fixed to all branch points $e_1, e_2, e_3$. In higher genus, there are 2g+1 branch points in the case that one of the 2g+2 point is fixed at $\infty$, while there are $\frac{g(g+1)}{2} \bullet \binom{2g+1}{g}$ number of constants $P_{IJ}(\frac{1}{2}\Omega_\delta)$. How can we relate these two, and how can we sum over spin structures, to have a final form of amplitudes in terms of symmetric functions of $e_1, e_2, .....e_{2g+1}$? These issues are clarified in the subsection 4.2, with a help of an old result of classical algebraic geometry. There is an explicit, simple way of determining $P_{IJ}(\frac{1}{2}\Omega_\delta)$, and it can be naturally regarded as a general tool of summing over spin structures for any genus in hyper elliptic case. The validity of this set of equations itself has nothing to do with the assumption we mentioned.

In our method, it is always convenient to fix one of the branch points at $\infty$, and use only 2g+1 points. In such a formulation, the genus g hyper elliptic string measure found in [5] should be written in a slightly modified form as:

$$\frac{1}{\sqrt{D}}[\text{product of differences of } e_i, \text{ depending on g number of chosen points out of } 2g+1]$$

(1.5)

where $\sqrt{D} \equiv \prod_{i<j}^{g}(e_i - e_j)$ is the square root of the discriminant of the curve.

For g=1, it is

$$\frac{(e_1 - e_3)}{\sqrt{D}}, \quad \frac{(e_3 - e_2)}{\sqrt{D}}, \quad \frac{(e_2 - e_1)}{\sqrt{D}}$$

For g>1, we will discuss using the measure found in [5], with being normalized as divided by $D$ in section 5. In section5, we see how the combinations of the biholomorphic 1 forms [1] in two loop four point function are reproduced by our method.

We will discuss the general structure of string amplitudes and future problems in section 6.

## 2. Review of the case of genus one

Throughout the document, $z_1, z_2 \ldots z_M$ are inserting points of vertex operators. In this chapter we define $E(z-w) = \dfrac{\theta_1(z-w)}{\dot\theta_1(0)}$ which means that in g=1 we do not use holomorphic 1 form and everything in this section is just a function. Szego kernel at g=1 is also written as a function here: $S_\delta(z-w) = \dfrac{\theta_{\delta+1}(z-w)}{\theta_{\delta+1}(0)E(z-w)}$ where $\delta$ represents a spin structure, $\delta = 1,2,3$. For the classical notations of theta functions, see Appendix A.

Our main interest for a moment is the product

$$S_\delta(z_1 - z_2)S_\delta(z_2 - z_3)\ldots S_\delta(z_M - z_{M+1}) \quad \text{with} \quad z_{M+1} = z_1 \qquad (2.1)$$

We will sometimes use the x variables just for convenience:

$$x_1 \equiv z_1 - z_2, \quad x_2 \equiv z_3 - z_2. \quad \ldots, x_M \equiv z_M - z_1 \qquad (2.2)$$

Obviously $\sum_{i=1}^{M} x_i = 0.$ (2.3)

We begin with a simple calculation starting from the Fay's trisecant identity:

$$\theta_{\delta+1}(z_1 + z_2 - w_1 - w_2)\vartheta_{\delta+1}(0)E(z_1 - z_2)E(w_1 - w_2)$$
$$= +\theta_{\delta+1}(z_1 - w_2)\vartheta_{\delta+1}(z_2 - w_1)E(z_1 - w_1)E(z_2 - w_2) - \theta_{\delta+1}(z_1 - w_1)\vartheta_{\delta+1}(z_2 - w_2)E(z_1 - w_2)E(z_2 - w_1)$$
(2.4)

Taking the derivative in $w_2$ and setting $w_2 = z_2$, we have

$$S_\delta(z_1 - z_2)S_\delta(z_2 - z_3) = -\frac{\partial_{z_1}\theta_{\delta+1}(z_1 - z_3)}{\theta_{\delta+1}(0)E(z_1 - z_3)} + S_\delta(z_1 - z_3)\partial_{z_2}\ln\frac{\theta_1(z_3 - z_2)}{\theta_1(z_1 - z_2)} \qquad (2.5)$$

The Szego kernel is related to the classical Weierstrass Pe function in g=1 as

$$S_\delta(z-w)^2 = P(z-w) - e_\delta \tag{2.6}$$

where $e_\delta$ is the value of the Pe function at half periods

$$e_\delta = P(\omega_\delta) \tag{2.7}$$

and this is the branch point of the curve.

Multiplying $S_\delta(z_3 - z_1)$ to eq. (2.5) and using eq. (2.6), we have

$$S_\delta(z_1 - z_2)S_\delta(z_2 - z_3)S_\delta(z_3 - z_1) = A + Be_\delta \tag{2.8}$$

where

$$A = +\frac{1}{2}\frac{\partial}{\partial z_1}P(z_3 - z_1) - P(z_3 - z_1)\{\frac{\partial}{\partial z_1}\ln\theta_1(z_1 - z_2) + \frac{\partial}{\partial z_2}\ln\theta_1(z_2 - z_3) + \frac{\partial}{\partial z_3}\ln\theta_1(z_3 - z_1)\} \tag{2.9}$$

$$B = \frac{\partial}{\partial z_1}\ln\theta_1(z_1 - z_2) + \frac{\partial}{\partial z_2}\ln\theta_1(z_2 - z_3) + \frac{\partial}{\partial z_3}\ln\theta_1(z_3 - z_1) \tag{2.10}$$

We do a little more: start from $S_\delta(z_2 - z_3)\, S_\delta(z_3 - z_1)$ and multiply $S_\delta(z_1 - z_2)$ and do the same calculation. Once again, start from $S_\delta(z_3 - z_1)\, S_\delta(z_1 - z_2)$ and multiply $S_\delta(z_2 - z_3)$ and do the same calculation. The all of these should give the same result. Sum all and divide by 3. The result is easily found to be, using the variables $x_1$, $x_2$, $x_3$ defined in (2.2),

$$S_\delta(z_1 - z_2)S_\delta(z_2 - z_3)S_\delta(z_3 - z_1) = S_\delta(x_1)S_\delta(x_2)S_\delta(x_3) = C + De_\delta \tag{2.11}$$

where

$$C = -\frac{1}{6}\{\sum_{i=1}^{3}\frac{\partial}{\partial x_i}P(x_i)\} - \frac{1}{3}\{\sum_{i=1}^{3}P(x_i)\}\cdot\{\sum_{i=1}^{3}\frac{\partial}{\partial x_i}\ln\theta_1(x_i)\} \tag{2.12}$$

$$D = \frac{1}{3}\sum_{i=1}^{3}\frac{\partial}{\partial x_i}\ln\theta_1(x_i) \tag{2.13}$$

One thing which should be noted is that the C and D are elliptic functions of vertex inserting points, written only by the differentials of log of theta-1 which are manifestly modular invariant, and have nothing to do with spin structures[1].

The spin structures of the 3 product of Szego kernel are factored out only in a constant $e_\delta$. We will see that this feature is not accidental at least in g=1, for any number of M.   The product of the Szego kernels always has this kind of decomposition as we will see in the generalized theorem .

---

[1] It looks that the order of the poles in D is one, but the order of any z is not 1, due to the condition $\sum_{i=1}^{3} x_i = 0$, and so the whole can be elliptic.

   Next, at first sight the form of C may look a little complex, but if we note an elementary identity:

$$\frac{1}{x_1 x_2 x_3} = \frac{1}{3}\{\frac{1}{x_1^3}+\frac{1}{x_2^3}+\frac{1}{x_3^3}\} - \frac{1}{3}(\frac{1}{x_1^2}+\frac{1}{x_2^2}+\frac{1}{x_3^2})(\frac{1}{x_1}+\frac{1}{x_2}+\frac{1}{x_3})$$

if   $x_1 + x_2 + x_3 = 0$ ,  (2.14)

and considering the pole of the derivative of the Pe function： $\frac{\partial}{\partial x_i}\frac{1}{x_i^2} = -\frac{2}{x_i^3}$,   then we can say that C is the elliptic function whose poles are of the simple form $\frac{1}{x_1 x_2 x_3}$ .

The   C   does not have poles such as $\frac{1}{x_1 x_2^2}$ , $\frac{1}{x_1 x_3^2}$ ,or $\frac{1}{x_3^3}$,   etc.
This feature of C is also not accidental.  Note that D is also such a type, it has symmetric poles of the form $\frac{1}{x_i}$ .  These simple pole structures will be real particle poles in the case of higher point amplitudes.

If we say  " elliptic function which has the pole  $\frac{1}{x_1 x_2 x_3}$ " , then such a function is uniquely determined by Liouville's theorem as in (2.12), though there may be plural ways of expressions.    This elliptic function can always be expressed by unique odd theta function on torus but the expression form may be long.   It is sometimes more convenient to attach a symbol,   say V ,   defined as follows, to such kind of function like C,  and use it instead of long theta function expressions.  This V itself, which shows simple pole structure which reflects the pole structure of the product of Szego kernels, can actually be used as a building block of elliptic function resulted from the contraction of fermion fields in the vertex operators, instead of its long theta function expressions.

**Definition:**     $_M V_K(x_1, x_2, x_3 ... x_M)$    is " the elliptic function whose poles are all symmetric combinations of the inverse of K number of variables out of total M numbers $x_1, x_2, .... x_M$".    That is, if we choose K number of variables $a_1, a_2, ... a_K$ out of M number of   $x_1, x_2, ... x_M$

$$_M V_K(x_1, x_2, x_3 ... x_M) = \frac{1}{number\ of\ combinations} \sum_{all\ combinations} function\ of\ the\ pole\ \frac{1}{a_1 a_2 .. a_K}$$

(2.15)

We always assume $\sum_{i=1}^{M} x_i = 0$.

Examples are

$$_3V_1(x_1, x_2, x_3) = \frac{1}{3}\sum \text{ elliptic function with the pole } \frac{1}{x_i} = \frac{1}{3}\sum_{i=1}^{3} \frac{\partial}{\partial x_i} \ln \vartheta_1(x_i) = D$$

$$_3V_3(x_1, x_2, x_3) = \text{elliptic function with the pole } \frac{1}{x_1 x_2 x_3} = C$$

$$_5V_1(x_1, x_2, x_3, x_4, x_5) = \frac{1}{5}\sum_{i=1}^{3} \frac{\partial}{\partial x_i} \ln \vartheta_1(x_i)$$

The last one appears in the five point amplitudes as we will see.

Also, as we will see below, there is a general formula for arbitrary M in which V is expressed as a ratio of two Pe function determinants. For example,

$$_3V_3(x_1, x_2, x_3) = C = \frac{\begin{vmatrix} P(x_1) & P'(x_1) \\ P(x_2) & P'(x_2) \end{vmatrix}}{\begin{vmatrix} 1 & P(x_1) \\ 1 & P(x_2) \end{vmatrix}}, \quad _3V_1(x_1, x_2, x_3) = D = \frac{\begin{vmatrix} 1 & P'(x_1) \\ 1 & P'(x_2) \end{vmatrix}}{\begin{vmatrix} 1 & P(x_1) \\ 1 & P(x_2) \end{vmatrix}} \qquad (2.16)$$

In this notation,
$$S_\delta(x_1)S_\delta(x_2)S_\delta(x_3) = {}_3V_3(x_1, x_2, x_3) + {}_3V_1(x_1, x_2, x_3)e_\delta \qquad (2.17)$$

One can always regard this V as a convenient, compact expression of elliptic function expressed by the unique odd theta function on torus, because the function itself is uniquely determined once the pole structure is given. The order of the pole is always 1 for any x in fermion contractions. But we again note that the order of any z is not 1, due to the condition $\sum_{i=1}^{M} x_i = 0$, and so the whole can be elliptic.

This M=3 case (2.17) can be proved without using Fay's formula as we will see in the proof of the theorem below. And starting from (2.17) we can derive Fay's formula, as will be noted at the end of Appendix A. This means that the formula (2.17) is equivalent to the Fay's formula for M=3.

It is not so straightforward to obtain the similar formulas in the cases of M=4, 5, 6... by only using Fay's formula of (3x3), (4x4)... step by step. By adopting a different method in which Fay's formula is not used, we can prove the following generalizations of the factorized formula eq. (2.17):

**Theorem:**

The following decomposition identity holds in the case of $\sum_{i=1}^{M} x_i = 0$:

$$\prod_{i=1}^{M} S_\delta(x_i) = \sum_{K=0}^{\left[\frac{M}{2}\right]} {}_M V_{M-2K}(x_1, x_2, ... x_M) \cdot (e_\delta)^K \quad (\text{for each of } \delta = 1, 2, 3) \quad (2.18)$$

under the same definition of ${}_M V_K(x_1, x_2, x_3 ... x_M)$ as in (2.15), and ${}_M V_0 \equiv 1$. The summation is until the integer which does not exceed $\frac{M}{2}$.

The ${}_M V_{M-2K}(x_1, x_2, ... x_M)$ has a manifestly elliptic form for arbitrary M and K, using the ratio of Pe functions. In the case that M is even,

$$
{}_M V_{M-2K} = \frac{\begin{vmatrix} 1 & P(x_1) & \cdots & P'(x_1)P^{K-2}(x_1) & P'(x_1)P^{K-1}(x_1) & \cdots & P^{\frac{M}{2}}(x_1) \\ 1 & P(x_2) & \cdots & P'(x_2)P^{K-2}(x_2) & P'(x_1)P^{K-1}(x_2) & \cdots & P^{\frac{M}{2}}(x_2) \\ \vdots & \vdots & & & & & \vdots \\ \vdots & \vdots & & & & & \vdots \\ 1 & P(x_{M-1}) & \cdots & P'(x_{M-1})P^{K-2}(x_{M-1}) & P'(x_{M-1})P^{K-1}(x_{M-1}) & \cdots & P^{\frac{M}{2}}(x_{M-1}) \end{vmatrix}}{\begin{vmatrix} 1 & P(x_1) & P'(x_1) & \cdots & P'(x_1)P^{\frac{M}{2}-2}(x_1) \\ 1 & P(x_2) & P'(x_2) & \cdots & P'(x_2)P^{\frac{M}{2}-2}(x_2) \\ \vdots & \vdots & \vdots & & \vdots \\ \vdots & \vdots & \vdots & & \vdots \\ 1 & P(x_{M-1}) & P'(x_{M-1}) & \cdots & P'(x_{M-1})P^{\frac{M}{2}-2}(x_{M-1}) \end{vmatrix}}
$$

(2.19)

In the determinants, the Pe function is lined up so that the order of poles are increasing as $P, P', P^2, P'P, P^3, P'P^2, \ldots$ until the pole is of order M. The derivative of Pe function contained is only up to the first order, $P'$, and all others are monomials of $P$. In the numerator, $P^K$ terms are replaced by 1 and are moved to the left end. In the denominator, $P^{\frac{M}{2}}$ terms are replaced by 1, and are moved to the left end.

In the case that M is odd, the components in the right end column in the numerator are replaced with

$$P'(x_1)P^{\frac{M-3}{2}}(x_1),\ P'(x_1)P^{\frac{M-3}{2}}(x_1),........P'(x_{M-1})P^{\frac{M-3}{2}}(x_{M-1})\ \ ,$$

and the components in the right end column in the denominator are replaced with

$$P^{\frac{M-1}{2}}(x_1),\ P^{\frac{M-1}{2}}(x_1),........P^{\frac{M-1}{2}}(x_{M-1})\ \ .$$

In both cases, the highest order of the pole is M in the numerator, and M-1 in the denominator.

The formula (2.19) is for $K \geq 2$. It is easy to write down the case for K=0 and K=1 but those are not used in the non-zero contributions after summing over spin structures.

Therefore, in multi point massless amplitudes in g=1, the product of Szego Kernels is always decomposed as in (1.4), which has all nice features already described. This is an example of the decomposition of the quantities which define positions of operators in the operator-product expansions and those related to the moduli of Riemann surfaces.

The formula (2.18) is natural in the following sense: the product of Szego kernel has obviously the pole of the first order for each of the variables $\frac{1}{x_1 x_2 .. x_M}$, but it is not elliptic. This is decomposed to the summation of manifestly modular invariant elliptic functions V whose poles are of the form $\frac{1}{a_1 a_2 .. a_K}$, where $a_1, a_2,...a_K$ are K number of variables chosen from $x_1, x_2,...x_M$. Each of the elliptic functions is multiplied by spin structure dependent function Q which only depends on the moduli of torus.

The formula (2.18) says that the spin structure dependence of the amplitude for any M is only on the constant $e_\delta$, that is, the Pe function value at half periods $P(\omega_\delta)$. After the summation, this constant part sometimes gives zero, sometimes constant values, and in general gives constant modular forms expressed as the fundamental symmetric polynomials of the three branch points by

$$e_1 + e_2 + e_3 = 0 \qquad e_1 e_2 + e_2 e_3 + e_3 e_1 = -\frac{g_2}{4} \qquad e_1 e_2 e_3 = \frac{g_3}{4}$$
(2.20)

We do not yet have the similar theorem for the higher genus case, and at present we will see the similar thing is true of M<4 cases for arbitrary g in the section3. This feature is expected to hold for any M, any g.

Proof of the theorem [2]

We start with the Frobenius – Stickelberger formula:

$$\begin{vmatrix} 1 & P(x_1) & P'(x_1) & \cdots & P^{(M-2)}(x_1) \\ 1 & P(x_2) & P'(x_2) & \cdots & P^{(M-2)}(x_2) \\ \vdots & \vdots & \vdots & & \vdots \\ \vdots & \vdots & \vdots & & \vdots \\ 1 & P(x_M) & P'(x_M) & \cdots & P^{(M-2)}(x_M) \end{vmatrix}$$

$$= (-1)^{\frac{(M-1)(M-2)}{2}} 1!2!\cdots(M-1)! \frac{\sigma(x_1+x_2+\cdots x_M)\prod_{\lambda<\mu}\sigma(x_\lambda - x_\mu)}{\prod_{k=1}^{M}\sigma^M(x_k)}$$

(2.21)

This is often denoted recently as a corollary of Fay's formula, but of course it is possible to prove this classical formula without using Fay's formula.

We assume M is even for a while. Since

$$P^{(2n)}(z) = \frac{d^{2n}P(z)}{dz^{2n}} = polynomial\ of\ P(z)\ of\ \deg ree\ n+1 \quad (A.3)$$

$$P^{(2n+1)}(z) = P'(z) * [polynomial\ of\ P(z)\ of\ \deg ree\ n] \quad (A.4)$$

eq.(2.21) is modified as

$$\begin{vmatrix} 1 & P(x_1) & P'(x_1) & P^2(x_1) & P'(x_1)P(x_1) & P^3(x_1)\ldots & P^{\frac{M}{2}}(x_1) \\ 1 & P(x_2) & P'(x_2) & P^2(x_2) & P'(x_2)P(x_2) & P^3(x_2)\ldots & P^{\frac{M}{2}}(x_2) \\ \vdots & \vdots & \vdots & & & & \vdots \\ \vdots & \vdots & \vdots & & & & \vdots \\ 1 & P(x_M) & P'(x_M) & P^2(x_M) & P'(x_M)P(x_M) & P^3(x_M)\ldots & P^{\frac{M}{2}}(x_M) \end{vmatrix}$$

$$= 2^{\frac{M-2}{2}} \frac{\sigma(x_1+x_2+\cdots x_M)\prod_{\lambda<\mu}\sigma(x_\lambda - x_\mu)}{\prod_{k=1}^{M}\sigma^M(x_k)} \quad (2.22)$$

so that each component of the determinant has only one term. The derivative of Pe function is only up to the first order and all components are monomials so as to make the investigation of the poles easy.

---

[2] We use a method in [23]. The logic in [23] is correct but to attach the desired meaning to the elliptic function V in its definition, we need to do slightly different argument starting from (2.22), not from (2.21). So the result (2.18) was not reported in [23] in a complete form.

If the sum of the variables are zero: $\sum_{i=1}^{M} x_i = 0$, which we note again that x are the difference of the inserting point of vertex operators z, then the right hand side of eq.(2.22), becomes zero. Therefore, for $z = x_1, x_2, \ldots x_M$, there exist $a_0, a_1, \ldots a_{M-2}$ which satisfy

$$1 + a_0 P(z) + a_1 P'(z) + a_2 P^2(z) + a_3 P'P(z) \cdots + a_{M-2} P^{\frac{M}{2}}(z) = 0 \tag{2.23}$$

Here $z$ is a new formal variable, and has nothing to do with the vertex inserting points. This is easily solved for $a_0, a_1, \ldots a_{M-2}$, whose explicit form we will use later.

Now we consider polynomials $f(x)$ and $h(x)$ defined as follows.

$$f(P(z)) \equiv \{h(P(x))\}^2 - \{a_1 P'(z) + a_3 P'(z)P(z) + a_5 P'(z)P^2(z) \cdots + a_{M-3} P'(z) P^{\frac{M}{2}-2}(z)\}^2 \tag{2.24}$$

$$h(P(z)) \equiv 1 + a_0 P(z) + a_2 P^2(z) + a_4 P^3(z) \cdots + a_{M-2} P^{\frac{M}{2}}(z) \tag{2.25}$$

The $f(x)$ is degree M polynomial, and can be easily written as :
$$f(P(z)) = h(P(z))^2 - \{P'(z)\}^2 [polynomial\ of\ P(z)] \tag{2.26}$$

On the other hand, if we factorize the definition of $f(x)$ in (2.24), then we can easily see that a equation of $f(x) = 0$ has M number of solutions at $x = P(x_1), P(x_2), \ldots P(x_M)$ by (2.23). Therefore, $f(x)$ can be written in a different way :

$$f(x) = a_{M-2}(x - P(x_1))(x - P(x_2))\ldots(x - P(x_M)) \tag{2.27}$$

If we compare $f(x)$ with the form of $\prod_{i=1}^{M} S_\delta(x_i)$, because of the relationship

$S_\delta(z)^2 = P(z) - e_\delta$ (2.6), the latter is found to be proportional to the square root of $f(x)$ with putting $x = e_\delta$. That is,

$$\prod_{i=1}^{M} S_\delta(x_i) = \frac{[f(e_\delta)]^{\frac{1}{2}}}{a_{M-2}} \tag{2.28}$$

If we note $P(\omega_\delta) = e_\delta$ then $f(e_\delta) = f(P(\omega_\delta))$, but since $P'(\omega_\delta) = 0$, we find from (2.26) that

$$f(e_\delta) = f(P(\omega_\delta)) = h(P(\omega_\delta))^2 - \{P'(\omega_\delta)\}^2 [polynomial\ of\ P(\omega_\delta)] = h(P(\omega_\delta))^2$$

(2.29)

Therefore the square root in (2.28) disappears, only leaving $h(e_\delta)$ :

$$\prod_{i=1}^{M} S_\delta(x_i) = \frac{[f(e_\delta)]^{1/2}}{a_{M-2}} = \frac{[f(P(\omega_\delta))]^{1/2}}{a_{M-2}} = \frac{h(e_\delta)}{a_{M-2}} \qquad (2.30)$$

Now solving (2.23) we have the explicit expressions of $a_0, a_1, \ldots a_{M-2}$ by Cramer's formula ( Note again that we have been assuming M is even after eq.(2.21) ) :

$$a_i = (-1) \frac{\begin{vmatrix} 1 & P(x_1) & \cdots & P'(x_1)P^{\frac{i}{2}-1}(x_1) & P'(x_1)P^{\frac{i}{2}}(x_1) & \cdots & P^{\frac{M}{2}}(x_1) \\ 1 & P(x_2) & \cdots & P'(x_1)P^{\frac{i}{2}-1}(x_2) & P'(x_1)P^{\frac{i}{2}}(x_2) & \cdots & P^{\frac{M}{2}}(x_2) \\ \vdots & \vdots & & & & & \vdots \\ \vdots & \vdots & & & & & \vdots \\ 1 & P(x_{M-1}) & \cdots & P'(x_1)P^{\frac{i}{2}-1}(x_{M-1}) & P'(x_1)P^{\frac{i}{2}}(x_{M-1}) & \cdots & P^{\frac{M}{2}}(x_{M-1}) \end{vmatrix}}{\begin{vmatrix} P(x_1) & P'(x_1) & \cdots & P^{\frac{M}{2}}(x_1) \\ P(x_2) & P'(x_2) & \cdots & P^{\frac{M}{2}}(x_2) \\ \vdots & \vdots & & \vdots \\ \vdots & \vdots & & \vdots \\ P(x_{M-1}) & P'(x_{M-1}) & \cdots & P^{\frac{M}{2}}(x_{M-1}) \end{vmatrix}}$$

(2.31)

for even i , and similar expression for odd i which will not be used. When we see (2.30) and explicit form of h(x) in (2.25), the dependence of $\prod_{i=1}^{M} S_\delta(x_i)$ on $x_i$ is through the ratio $\dfrac{a_j}{a_{M-2}}$ where j is even. The Pe function part in (2.25) becomes functions of constant $e_\delta$ when we calculate $f(e_\delta)$ .

Since both of denominator and numerator in the factor $\dfrac{a_j}{a_{M-2}}$ are manifestly elliptic because all are written by Pe function, the whole is also elliptic, and it is determined by its pole structures only. If we denote the inverse of the poles of Pe functions as y, it is apparent from (2.31) that what we should investigate is the ratio of the following two types of determinants :

$$\begin{vmatrix} 1 & y_1^2 & y_1^3 & \cdots & & \cdots & y_1^n \\ 1 & y_2^2 & y_2^3 & \cdots & & \cdots & y_2^n \\ \vdots & \vdots & \vdots & & & & \vdots \\ \vdots & \vdots & \vdots & & & & \vdots \\ 1 & y_n^2 & y_n^3 & \cdots & & \cdots & y_n^n \end{vmatrix} \text{ and } \begin{vmatrix} 1 & y_1^2 & y_1^3 & \cdots & y_1^{k-1} & y_1^{k+1} & \cdots & y_1^{n+1} \\ 1 & y_2^2 & y_2^3 & \cdots & y_2^{k-1} & y_2^{k+1} & \cdots & y_2^{n+1} \\ \vdots & \vdots & \vdots & & \vdots & \vdots & & \vdots \\ \vdots & \vdots & \vdots & & \vdots & \vdots & & \vdots \\ 1 & y_n^2 & y_n^3 & \cdots & y_n^{k-1} & y_n^{k+1} & \cdots & y_n^{n+1} \end{vmatrix}, \quad (2.32)$$

both of which are slightly different from Van der Monde type.

In Appendix B , we proved by this pole investigation that, up to over all constants,

$$\frac{a_j}{a_{M-2}} = {}_M V_{M-2-j}(x_1, x_2, x_3 \ldots x_M) \tag{2.33}$$

$$\frac{1}{a_{M-2}} = {}_M V_M(x_1, x_2, x_3 \ldots x_M) \tag{2.34}$$

That is, $\dfrac{a_j}{a_{M-2}}$ is an elliptic function whose pole is the symmetric sum of the inverse of $M-2-j$ number of products out of M number of variables $x_1, x_2, x_3 \ldots x_M$ .

Also, $\dfrac{1}{a_{M-2}}$ is an elliptic function whose pole is the inverse of product of all variables

$x_1, x_2, x_3 \ldots x_M$ .

It is straightforward to repeat the same arguments for odd M, and inserting this result into (2.30), we immediately have

$$\prod_{i=1}^{M} S_\delta(x_i) = \sum_{K=0}^{[\frac{M}{2}]} {}_M V_{M-2K}(x_1, x_2, \ldots x_M) \cdot e_\delta^K \qquad (\delta = 1, 2, 3) \tag{2.18}$$

## 3. Pole structures of M point massless boson amplitudes at g = 1 for arbitrary M

When we calculate M point massless amplitudes, one of the contractions of the vertex operators gives the following spin structure summation of the M products of Szego kernels:

$$\sum_{\delta=1,2,3} (-1)^\delta \left[ \frac{\theta_{\delta+1}(0|\tau)}{\dot{\theta}_1(0|\tau)} \right]^4 \prod_{i=1}^{M} S_\delta(x_i) \quad , \text{ which we denote as } G_M.$$

Note again that $x_i$ are the difference of points of vertex inserting points,

$$x_1 \equiv z_1 - z_2, \quad x_2 \equiv z_3 - z_2. \quad ...., \quad x_M \equiv z_M - z_1 \qquad (2.2)$$

By the well known formulas, $\dot{\theta}_1(0) = \pi \theta_2(0)\theta_3(0)\theta_4(0)$ (A.16) and Thomae type formulas (A.10), $G_M$ is written to be

$$G_M = \frac{1}{\sqrt{D}}\left[(e_1 - e_3)\prod_{i=1}^{M} S_2(x_i) + (e_3 - e_2)\prod_{i=1}^{M} S_1(x_i) + (e_2 - e_1)\prod_{i=1}^{M} S_3(x_i)\right] \qquad (3.1)$$

Here D is the discriminant of the curve, and we use its square root in the denominator:

$$\sqrt{D} = (e_1 - e_3)(e_3 - e_2)(e_2 - e_1) \qquad (3.2)$$

$$\prod_{i=1}^{M}(P(x_i) - e_\delta)^{1/2} = \prod_{i=1}^{M} S_\delta(x_i) \qquad (3.3)$$

This means that we take the superstring measure at g=1 as $\frac{1}{\sqrt{D}}(e_i - e_j)$ with appropriate GSO projection, which correspond to choosing one of the branch points out of 1, 2, ...2g+2 fixed at $\infty$. For our purpose, this is more convenient, rather than writing the measure choosing 2 points out of 4 points.

From (2.18) and (3.1), the general result of spin structure summation of M-point one loop amplitude is

$$G_M = \sum_{K=0}^{\left[\frac{M}{2}\right]} {}_M V_{M-2K} \cdot \frac{(e_1 - e_3)e_2^K + (e_3 - e_2)e_1^K + (e_2 - e_1)e_3^K}{(e_1 - e_3)(e_3 - e_2)(e_2 - e_1)} \qquad (3.4)$$

where the elliptic function V is given by (2.19) which has simple pole structures and can be expressed by unique odd theta function. The factor $\frac{(e_1 - e_3)e_2^K + (e_3 - e_2)e_1^K + (e_2 - e_1)e_3^K}{(e_1 - e_3)(e_3 - e_2)(e_2 - e_1)}$ is expressed by fundamental symmetric polynomials of $e_1, e_2, e_3$ for arbitrary value of K because

1) If we regard the factor as the function of $e_1$, then at $e_1 = e_2$ and at $e_1 = e_3$ the numerator becomes zero. Therefore the numerator contains the denominator, which means the whole is a polynomial.
2) It is manifestly symmetric.

Then, by $e_1 + e_2 + e_3 = 0$ $\quad e_1 e_2 + e_2 e_3 + e_3 e_1 = -\frac{g_2}{4}$ $\quad e_1 e_2 e_3 = \frac{g_3}{4}$ , this factor is always expressive by the modular forms $g_2, g_3$ only, which are proportional to Eisenstein series . It is easy to see that the value of K in (3.4), which is the degree of $e_\delta$ multiplied to the measure in the numerator, should be 2 or more than 2 to give non-zero result. We will see this critical number of degree is 2g in the general case of genus g. Also, as exemplified above, the theorem gives infinitely many theta identities to express the final form of the function of the vertex inserting points only by theta 1, or more precisely, the differentials of log of sigma function in genus one.

We can say the procedure above gives the complete patterns of pole structures of the M point amplitudes after spin structure summations. This also gives a natural proof of modular invariance of the M point amplitudes at g=1 level for any M.

We can do all summations only algebraically, and do not use Riemann quadratic relations. The Riemann identity is implicitly included in the modification of the whole of the fermionic contraction part using $\sum_{\delta=1,2,3}(-1)^\delta \left[\frac{\theta_{\delta+1}(0|\tau)}{\dot{\theta}_1(0|\tau)}\right]^4 = 0$, and partly in the theorem itself. In particular, since we can always have explicit modular invariant forms in V as for the operator inserting points dependent part by the theorem, the process of summing over spin structure is simpler than the standard methods.

Let us see some examples of spin structure summation and see some theta identities.
First, for the cosmological constant, in which there are no Szego kernels, the numerator of eq.(3.1) is trivially zero:
$$(e_1 - e_3) + (e_3 - e_2) + (e_2 - e_1) = 0$$
For the two and three point functions, we use

$$S_\delta(x_1)S_\delta(x_2) =\,_2V_2(x_1, x_2) \cdot + e_\delta \qquad x_1 = z_2 - z_1, \quad x_2 = z_1 - z_2 \qquad (3.5)$$

$$S_\delta(x_1)S_\delta(x_2)S_\delta(x_3) =\,_3V_3(x_1, x_2, x_3) \cdot +\,_3V_1(x_1, x_2, x_3)e_\delta \qquad (3.6)$$

$$x_1 = z_2 - z_1, \quad x_2 = z_3 - z_2, \quad x_3 = z_3 - z_1$$

All elliptic functions V have explicit forms written in unique odd theta function, but all are multiplied degree 1 polynomial of $e_\delta$ in the expression of $S_\delta S_\delta .... S_\delta$, therefore the spin structure summation is zero by a trivial relationship:

$$(e_1 - e_3)e_2 + (e_3 - e_2)e_1 + (e_2 - e_1)e_3 = 0$$

At the four point level, we have degree 2 polynomial of $e_\delta$,

$$S_\delta(x_1)S_\delta(x_2)S_\delta(x_3)S_\delta(x_4) = {}_4V_4(x_1,x_2,x_3,x_4) + {}_4V_2(x_1,x_2,x_3,x_4)e_\delta + {}_4V_0(x_1,x_2,x_3,x_4)e_\delta^2$$

(3.7)

$${}_4V_0(x_1,x_2,x_3,x_4) = 1.$$

Now there appeared the first non-zero result, only from $e_\delta^2$ in the last constant term due to a simple algebraic identity again:

$$\frac{(e_1 - e_3)e_2^2 + (e_3 - e_2)e_1^2 + (e_2 - e_1)e_3^2}{(e_1 - e_3)(e_3 - e_2)(e_2 - e_1)} = 1$$

From the fermion field contractions there appeared no elliptic functions V at this level, only gives a constant.

In calculating four-point function, there is one more Wick contraction term $S_\delta(z_2 - z_1)^2 S_\delta(z_3 - z_4)^2$ which also gives degree 2 polynomial of $e_\delta$ by eq. (2.6).

At the five point level, we have a contraction of the form:

$$S_\delta(x_1)S_\delta(x_2)S_\delta(x_3)S_\delta(x_4)S_\delta(x_5) = {}_5V_5 + {}_5V_3\, e_\delta + {}_5V_1\, e_\delta^2 \qquad (3.8)$$

After spin structure summation, only the term proportional to ${}_5V_1\, e_\delta^2$ survives, giving a term proportional to elliptic function ${}_5V_1$.

By the theorem in the last section, ${}_5V_1$ is the elliptic function whose poles are $\sum_{i=1}^{5} \frac{1}{x_i}$, or $\sum_{i=1}^{5} \frac{1}{z_{i+1} - z_i}$, $z_6 = z_1$. As already noted and as shown in the proof of the theorem, the V are always expressed by the ratio of two determinants of Pe functions, but it is sometimes convenient to rewrite them in terms of unique odd theta function on torus. This pole structure matches $\sum_{i=1}^{5} \varsigma(x_i)$, where $\varsigma$ is the Weierstrass $\varsigma$ function. Note that each $\varsigma$ function is not elliptic, but the summation of all these five terms is elliptic, because shifting all the variables by period cancels the redundant terms due to the condition $\sum_{i=1}^{5} x_i = 0$ .

The $\varsigma$ is related to odd theta function as $\sum_{i=1}^{5} \varsigma(x_i) = \sum_{i=1}^{5} \frac{\partial}{\partial x_i} \ln \theta_1(x_i|\tau)$, therefore we conclude that $_5V_1 = \sum_{i=1}^{5} \frac{\partial}{\partial x_i} \ln \theta_1(x_i|\tau)$ ,which is also proportional to

$$\begin{vmatrix} 1 & P(x_1) & P'(x_1) & P^{(3)}(x_1) \\ 1 & P(x_2) & P'(x_2) & P^{(3)}(x_2) \\ 1 & P(x_3) & P'(x_3) & P^{(3)}(x_3) \\ 1 & P(x_4) & P'(x_4) & P^{(3)}(x_4) \end{vmatrix} \bigg/ \begin{vmatrix} 1 & P(x_1) & P'(x_1) & P^{(2)}(x_1) \\ 1 & P(x_2) & P'(x_2) & P^{(2)}(x_2) \\ 1 & P(x_3) & P'(x_3) & P^{(2)}(x_3) \\ 1 & P(x_4) & P'(x_4) & P^{(2)}(x_4) \end{vmatrix} . \tag{3.9}$$

under the condition $\sum_{i=1}^{5} x_i = 0$.  See (B.6) in Appendix B.  This means that by the theorem we get a theta identity which is necessary to sum over five point amplitudes:

$$G_5 \equiv \sum_{\delta=1,2,3} (-1)^{\nu} \left[ \frac{\theta_{\delta+1}(0|\tau)}{\theta_1'(0|\tau)} \right]^4 \prod_{i=1}^{5} S_\delta(x_i) = \sum_{i=1}^{5} \frac{\partial}{\partial x_i} \ln \theta_1(x_i|\tau) \tag{3.10}$$

This modular invariant theta function will be included in the final form of the amplitude after summing over spin structures.

As one more example, consider a Wick contraction of the six point calculation.

$$S_\delta(x_1)S_\delta(x_2)S_\delta(x_3)S_\delta(x_4)S_\delta(x_5)S_\delta(x_6) = {}_6V_6 + {}_6V_4\, e_\delta + {}_6V_2\, e_\delta^2 + {}_6V_0\, e_\delta^3 \tag{3.11}$$

As for the degree-3 polynomial of $e_\delta$, we have

$$\frac{(e_1 - e_3)e_2^3 + (e_3 - e_2)e_1^3 + (e_2 - e_1)e_3^3}{(e_1 - e_3)(e_3 - e_2)(e_2 - e_1)} = e_1 + e_2 + e_3 = 0,$$

Then only the term proportional to $_6V_2\, e_\delta^2$ survives, giving a term proportional to $_6V_2$.

The $_6V_2$ is the elliptic function whose poles are $\sum_{i,j=1}^{6} \frac{1}{x_i \cdot x_j}$ where $i, j$ are different each other.  A way to express this by elliptic function is

$$_6V_2 = \frac{1}{2}[\sum_{i=1}^{6} \varsigma(x_i|\tau)]^2 + \frac{1}{2}[\sum_{i=1}^{6} P(x_i|\tau)] \text{ , which equals to}$$

$$= \frac{1}{2}[\sum_{i=1}^{6} \frac{\partial}{\partial x_i} \ln \theta_1(x_i|\tau)]^2 + \frac{1}{2}[\sum_{i=1}^{6} \frac{\partial^2}{\partial x_i^2} \ln \theta_1(x_i|\tau) + 12\eta_1] \tag{3.12}$$

The first term of the right hand is the elliptic function with poles $\frac{1}{x_i \cdot x_j}$, and the second term is just added so as to subtract elliptic functions with poles of $i = j$.

The theorem means that we get a theta identity which is necessary to sum over six point amplitudes:

$$G_6 \equiv \sum_{\delta=1,2,3} (-1)^\nu \left[\frac{\theta_{\delta+1}(0|\tau)}{\theta_1'(0|\tau)}\right]^4 \prod_{i=1}^{6} S_\delta(x_i) = \sum_{i=1}^{6} \frac{1}{2}[\frac{\partial}{\partial x_i}\ln\theta_1(x_i|\tau)]^2 + \frac{1}{2}\left[\sum_{i=1}^{6}\frac{\partial^2}{\partial x_i^2}\ln\theta_1(x_i|\tau) + 12\eta_1\right]$$

(3.13)

The additional term $12\eta_1$ from Pe function is interpreted that it comes from the regularization of Green functions in ghost pyramid formalism[27].

Since we saw that the terms for K=0, 1, 3 in (3.4) is zero, we can write

$$G_M = {}_M V_{M-4} + \sum_{K=4}^{\left[\frac{M}{2}\right]} {}_M V_{M-2K} \cdot \frac{(e_1-e_3)e_2^K + (e_3-e_2)e_1^K + (e_2-e_1)e_3^K}{(e_1-e_3)(e_3-e_2)(e_2-e_1)}$$

This means that for M<8, only the first term contributes to the amplitudes. In particular, for the four point amplitudes, the first term is unity.

## 4. Higher genus

In higher genus, the disconnected parts of string amplitudes will depend on the spin structures through the simple product of M Szego kernels $\prod_{i=1}^{M} S_\delta(z_i, z_{i+1})$ with the condition $z_{i+1} = z_1$, except the string measure.

This product has a pole $\frac{1}{x_1 x_2 ... x_M}$, where $x_i$ are conventional expression of the difference of operator inserting points, (2.2). Choose a subset of $x_i$, $a_1, a_2, ... a_K$, $K \leq M$. In general, the final form of the amplitude after summing over spin structure will be expressed by a combination of modular invariant functions with the pole $\frac{1}{a_1 a_2 .. a_K}$ for possible Ks and constant modular forms, in analogy with the case of g=1. The form of the modular invariant function of operator inserting points will be uniquely determined by the pole $\dfrac{1}{a_1 a_2 .. a_K}$ . This function will be described by unique theta function which has good property under the modular transformations, as the unique odd theta function played such role on the torus. Can any of the disconnected string amplitudes have different type of structures? We expect that such unique theta function is differentials of higher genus sigma function as was the case in g=1, as will be explained below.

4-1  Some calculations

In Appendix A of [1], some variants are computed starting from Fay's trisecant identity in genus g:

$$\theta[\delta](z_1 + z_2 - w_1 - w_2)\theta[\delta](0)E(z_1,z_2)E(w_1,w_2)$$
$$= +\theta[\delta](z_1 - w_2)\theta[\delta](z_2 - w_1)E(z_1,w_1)E(z_2,w_2) - \theta[\delta](z_1 - w_1)\theta[\delta](z_2 - w_2)E(z_1,w_2)E(z_2,w_1)$$
(4.1)

Two of them, obtained by taking the derivative in $w_2$ and setting $w_2 = z_2$ , are

$$S_\delta(z_1,z_2)S_\delta(z_2,z_3) = -\omega_I(z_2)\frac{\partial_I \theta[\delta](z_1-z_3)}{\theta[\delta](0)E(z_1,z_3)} + S_\delta(z_1,z_3)\partial_{z_2}\ln\frac{E(z_3,z_2)}{E(z_1,z_2)}$$
(4.2)

$$S_\delta(z,w)S_\delta(w,z) = -[S_\delta(z,w)]^2 = -\partial_z\partial_w \ln E(z,w) - \omega_I(z)\omega_J(z_2)\frac{\partial_I\partial_J\theta[\delta](0)}{\theta[\delta](0)}$$
(4.3)

where

$S_\delta(z,w) = \dfrac{\theta[\delta](z-w)}{\theta[\delta](0)E(z,w)}$ is the Szego kernel, $E(z,w) = \dfrac{\theta[\nu](z-w,\Omega)}{h_\nu(z)h_\nu(w)}$ is the Prime form in the standard notations.

In the following, we adopt one odd theta function with the suffix $\nu$ . The Prime form does not depend on the choice of $\nu$ , but later we will discuss about this index in theta function.

Multiplying $h_\nu(z_1)h_\nu(z_2) h_\nu(w_1)h_\nu(w_2)$ on the both side of Fay's formula, we repeat the same calculation as that of deriving (4.2),(4.3). Noting that $\partial_{w_2}\theta[\nu](z_2 - w_2) \to -h_\nu(z_2)^2$ when $w_2 \to z_2$ , and multiplying $h_\nu(z_1)h_\nu(z_3)$ , we have

$$S_\delta(z_1,z_2)S_\delta(z_2,z_3) = -\omega_I(z_2)\frac{\partial_I \theta[\delta](z_1-z_3)}{\theta[\delta](0)E(z_1,z_3)} + S_\delta(z_1,z_3)\partial_{z_2}\ln\frac{\theta[\nu](z_3-z_2)}{\theta[\nu](z_1-z_2)}$$

(4.4)

That is, the Prime form in the log of eq. (4.2) is replaced with theta function, and nothing else changed.

Taking the limit $w_3 \to z_1$, we also have

$$S_\delta(z,w)S_\delta(w,z) = -\partial_z\partial_w \ln \vartheta[\nu](z-w) - \omega_I(z)\omega_J(w)\frac{\partial_I\partial_J\theta[\delta](0)}{\theta[\delta](0)}$$

$$= \sum_{I,J=1}^g (\partial_I\partial_J \ln \vartheta[\nu](z-w) - \frac{\partial_I\partial_J\theta[\delta](0)}{\theta[\delta](0)})\omega_I(z)\omega_J(w) \quad (4.5)$$

Multiplying $S_\delta(z_3,z_1)$ to (4.4) and using (4.5) as well as $\partial_I h_\nu(z) = 0 = \partial_I \omega(z)$, we have

$$S_\delta(z_1,z_2)S_\delta(z_2,z_3)S_\delta(z_3,z_1) = [\,\{-\partial_J\partial_K \ln \theta[\nu](z_3-z_1) + \frac{\partial_I\partial_J\theta[\delta](0)}{\theta[\delta](0)}\}\partial_I L$$

$$+ \frac{1}{2}\partial_I\partial_J\partial_K \ln \theta[\nu](z_3-z_1)\,]\omega_I(z_2)\omega_J(z_3)\omega_K(z_1)$$

(4.6)

where we defined

$$L \equiv \ln \theta[\nu](z_1-z_2) + \ln \theta[\nu](z_2-z_3) + \ln \theta[\nu](z_3-z_1) \quad (4.7)$$

Doing the same calculation for $S_\delta(z_2,z_3)S_\delta(z_3,z_1)S_\delta(z_1,z_2)$ and $S_\delta(z_3,z_1)S_\delta(z_1,z_2)S_\delta(z_2,z_3)$ which are all the same, sum them and divided by 3, we have

$$S_\delta(z_1,z_2)S_\delta(z_2,z_3)S_\delta(z_3,z_1) = \sum_{I,J,K=1}^g [\,A^1{}_{IJK} + A^2{}_{IJK} + A^3{}_{IJK}\,]\omega_I(z_2)\omega_J(z_3)\omega_K(z_1)$$

(4.8)

in which

$$A^1{}_{IJK} = \frac{1}{3}[\frac{\partial_J\partial_K\theta[\delta](0)}{\theta[\delta](0)}\partial_I L + \frac{\partial_K\partial_I\theta[\delta](0)}{\theta[\delta](0)}\partial_J L + \frac{\partial_I\partial_J\theta[\delta](0)}{\theta[\delta](0)}\partial_K L] \quad (4.9)$$

$$A^2{}_{IJK} = -\frac{1}{3}[\,\partial_J\partial_K \ln \theta[\nu](z_1-z_2)\cdot\partial_I L + \partial_I\partial_K \ln \theta[\nu](z_2-z_3)\cdot\partial_J L + \partial_I\partial_J \ln \theta[\nu](z_3-z_1)\cdot\partial_K L\,]$$

(4.10)

$$A^3{}_{IJK} = \frac{1}{6}[\,\partial_I\partial_J\partial_K L\,] \quad (4.11)$$

Introducing another type of constants, periods of second kind Abelian differentials $r_J$ integrated around $A_I$ cycles [ For the notations, See Appendix C ]:

$$\eta_{IJ} = \oint_{A_I} r_J \qquad (C.3)$$

Define
$$P_{IJ}(\frac{1}{2}\Omega_\delta) \equiv \frac{1}{2}\eta_{IJ} - \frac{\partial_I \partial_J \theta[\delta](0)}{\theta[\delta](0)} \qquad (4.12)$$

The fundamental object is the higher genus sigma function, defined after adopting an appropriate theta function, schematically written as

$$\sigma(u) = c \exp(-\frac{1}{2}\sum_{I,J=1}^{g} u_I \eta_{IJ} u_J) \vartheta\begin{bmatrix}a\\b\end{bmatrix}(u) \qquad (4.13)$$

where the choice of the index $a, b$ is discussed later in the hyper elliptic case. (See (4.35).) We do not mention the over-all constant c. From this expression the higher genus Pe functions are defined:

$$P_{IJ}(u) = -\frac{\partial^2}{\partial u_I \partial u_J} \ln \sigma(u) \qquad (4.14)$$

The constants $P_{IJ}(\frac{1}{2}\Omega_\delta)$ means the Pe function values at the summation of selected g number of half periods corresponding to one even spin structure, as we will see in the next subsection. This is a natural extension of Weierstrass' Pe function values at "half-period" points of genus one Riemann surface, $P(\omega_\delta)$, that is, the constants $e_\delta$.

This setting can be done rather in a general way, but we will describe in detail only in hyper elliptic in the following subsection. The final goal is to express the fermion contraction part of the amplitude by the manifestly modular invariant differentials of sigma function and polynomials of $P_{IJ}(\frac{1}{2}\Omega_\delta)$ in analogy of the g=1 result. At present this can be done only partially in the higher genus case up to M=3.

It is straightforward to write (4.5) (4.8) in terms of the constant $P_{IJ}(\frac{1}{2}\Omega_\delta)$ :

$$S_\delta(z_1, z_2)S_\delta(z_2, z_1) = \sum_{I,J=1}^{g} [\partial_I \partial_J \ln \vartheta[\nu](z_1 - z_2) - \frac{1}{2}\eta_{IJ} + P_{IJ}(\frac{1}{2}\Omega_\delta)] \omega_I(z_1)\omega_J(z_2)$$

$$. \qquad (4.15)$$

$$S_\delta(z_1, z_2)S_\delta(z_2, z_3)S_\delta(z_3, z_1) = [-\{\partial_J \partial_K \ln \vartheta[\nu](z_3 - z_1) - \frac{1}{2}\eta_{IJ} + P_{IJ}(\frac{1}{2}\Omega_\delta)\}\partial_I L$$

$$+ \frac{1}{2}\partial_I \partial_J \partial_K \ln \vartheta[\nu](z_3 - z_1)]\omega_I(z_2)\omega_J(z_3)\omega_K(z_1) \qquad (4.16)$$

Or, in a symmetric form,

$$S_\delta(z_1,z_2)S_\delta(z_2,z_3)S_\delta(z_3,z_1) = \sum_{I,J,K=1}^{g} [B^1{}_{IJK} + B^2{}_{IJK} + B^3{}_{IJK}]\omega_I(z_2)\omega_J(z_3)\omega_K(z_1)$$

(4.17)

where

$$B^1{}_{IJK} = -\frac{1}{3}[\;P_{JK}(\frac{1}{2}\Omega_\delta)\,\partial_I L + P_{KI}(\frac{1}{2}\Omega_\delta)\,\partial_J L + P_{IJ}(\frac{1}{2}\Omega_\delta)\,\partial_K L] \quad (4.18)$$

$$B^2{}_{IJK} = +\frac{1}{3}\{[\partial_J\partial_K \ln \vartheta[\nu](z_1-z_2) - \frac{1}{2}\eta_{JK}]\cdot \partial_I L + [\partial_I\partial_K \ln \vartheta[\nu](z_2-z_3) - \frac{1}{2}\eta_{IK}]\,\partial_J L$$
$$+ [\partial_I\partial_J \ln \vartheta[\nu](z_3-z_1) - \frac{1}{2}\eta_{IJ}]\cdot \partial_K L\}$$

(4.19)

$$B^3{}_{IJK} = -\frac{1}{6}[\,\partial_I\partial_J\partial_K L] \qquad (4.20)$$

The period $\eta_{IJ}$ will be independent of the spin structures. It is not contained in the first derivative term L in general. The $\eta_{IJ}$ always appears in the second derivative terms of theta functions since it comes from Pe functions.

Now up to 3 point functions their spin structure dependence is obvious. The constant $P_{IJ}(\frac{1}{2}\Omega_\delta)$ , which is the only one object related to the spin structures up to this level, is 'factorized' as degree 0 or 1 polynomial in the expressions.

In the 3 point function, the $B^2{}_{IJK} + B^3{}_{IJK}$ has nothing to do with the spin structure, has the pole $\dfrac{1}{(z_1-z_2)(z_2-z_3)(z_3-z_1)}$ , and this will be a modular invariant theta function after rewriting it in sigma function. This is a generalization of ${}_3V_3$ in the g=1 case, and will be almost uniquely determined once the pole is given. After the summation over spin structure this part will be zero by the same reason that the cosmological constant vanishes.

The $B^1{}_{IJK}$ has the linear dependence on $P_{IJ}(\frac{1}{2}\Omega_\delta)$, and the spin structure summation can be done independently from the function ${}_3V_1^g$ or $\partial_I L$, $\partial_J L$, $\partial_K L$ multiplied.

Unfortunately it is not straightforward to obtain the results for M>3 by repeating the calculations in this way using Fay's formula. This was already the case for g=1; in g=1, another idea of using Frobenius – Stickelberger formula was necessary. In higher genus, the same idea may be useful to obtain the decomposed formula consecutively for M>3. If this point is successfully solved, it will give one of few methods of calculating higher point super string amplitudes in higher genus systematically. By using the results of M=0,2,3, we discuss the possible form of g loop 4 point amplitudes in section6.

4-2  The hyper elliptic case

In this subsection, the following facts will be shown: In hyper elliptic case, $\frac{1}{2}\Omega_\delta$ is shown to be equal to the summation of half periods corresponding to the number of g branch points $x_1, x_2, ....x_g$ chosen from the total 2g+1 points $e_1, e_2, .....e_{2g+1}$.[3] The fact that one spin structure corresponds to one such choice of g number of branch points is well known. We will connect it to the generalized Pe function values, by considering the summation of half periods of such chosen g points. Further, by applying a theorem obtained in the 19th century, we can have $\frac{g(g+1)}{2}$ number of linear equations by which $P_{IJ}(\frac{1}{2}\Omega_\delta)$ is solved as a function of $x_1, x_2, ....x_g$. The possibility of applying the theorem in this way is the main reason of adopting sigma function approach. This solution will allows us to sum over the possible $x_1, x_2, ....x_g$ only algebraically.

Once some settings on the Riemann surface are given, it is straightforward to see the exact meaning of the symbol $P_{IJ}(\frac{1}{2}\Omega_\delta)$ and how it is related to the constant $\frac{\partial_I \partial_J \theta[\delta](0)}{\theta[\delta](0)}$ which appeared in the calculations above. First we explain some definitions along [24]. See also [30][31].

For each of the branch points $e_j$, define a vector $U_j$ for each normalized holomorphic 1-forms $\omega_i$, as

$$U^i_j = \int_\infty^{e_j} \omega_i \equiv E^i_{0j} + \Omega E^i_{1j} \tag{4.21}$$

where vectors $E_{0j}$ and $E_{1j}$ are also defined in this expression, and modulo 1 all components of the vectors $E_{0j}$ and $E_{1j}$ are either 0 or 1/2. Defining $2\times g$ matrices

$$[U_j] = \begin{bmatrix} E^t_{1j} \\ E^t_{0j} \end{bmatrix} = \begin{bmatrix} E_{1j_1}, E_{1j_2}...E_{1j_g} \\ E_{0j_1}, E_{0j_2}...E_{0j_g} \end{bmatrix} \tag{4.22}$$

which are to be a basis for the characteristics of the theta function, we will determine the components of these matrix explicitly.

---

[3] The $x_1, x_2, ....x_g$ are new variables, not the one defined in (2.2).

Start from the obvious form, $[U_{2g+2}] = [0]$. Using the notation $f_k = \frac{1}{2}(\delta_{1k}, \delta_{2k},....\delta_{gk})$ where $\delta_{ij}$ is the Kronecker symbol, and $\Omega_k$ for the k-th column vector of the period matrix, we find

$$U^i_{2g+1} = U^i_{2g+2} - \sum_{k=1}^{g} \int_{e_{2k-1}}^{e_{2k}} \omega_i = \sum_{k=1}^{g} f_k \qquad \therefore [U_{2g+1}] = \frac{1}{2}\begin{bmatrix} 0 & 0.......0 & 0 \\ 1 & 1.......1 & 1 \end{bmatrix} \tag{4.23}$$

$$U^i_{2g} = U^i_{2g+1} - \int_{e_{2g+1}}^{e_{2g}} \omega_i = \sum_{k=1}^{g} f_k + \Omega_g \qquad \therefore [U_{2g}] = \frac{1}{2}\begin{bmatrix} 0 & 0.......0 & 1 \\ 1 & 1.......1 & 1 \end{bmatrix} \tag{4.24}$$

$$U^i_{2g-1} = U^i_{2g} - \int_{e_{2g}}^{e_{2g-1}} \omega_i = \sum_{k=1}^{g-1} f_k + \Omega_g \qquad \therefore [U_{2g}] = \frac{1}{2}\begin{bmatrix} 0 & 0.......0 & 1 \\ 1 & 1.......1 & 0 \end{bmatrix} \tag{4.25}$$

Then, in general,
$$[U_{2k+2}] = \frac{1}{2}\begin{bmatrix} k\text{ number of } 0 & 1\,0...0 \\ k\text{ number of } 1 & 1\,0...0 \end{bmatrix} \tag{4.26}$$

$$[U_{2k+1}] = \frac{1}{2}\begin{bmatrix} k\text{ number of } 0 & 1\,0...0 \\ k\text{ number of } 1 & 0\,0...0 \end{bmatrix} \tag{4.27}$$

and
$$[U_2] = \frac{1}{2}\begin{bmatrix} 1\,0...0 \\ 1\,0...0 \end{bmatrix} \tag{4.28} \qquad [U_1] = \frac{1}{2}\begin{bmatrix} 1\,0...0 \\ 0\,0...0 \end{bmatrix} \tag{4.29}$$

For the points $e_{2n}$, the characteristics with $n = 1,...g$ are odd, with others are even, except that for $[U_{2g+2}]$ is zero.

It is known that the vector of Riemann constants has the form

$$\Delta^i = -\sum_{k=1}^{g} \int_{\infty}^{e_{2k}} \omega_i = -\sum_{k=1}^{g} U^i_{2k} \tag{4.30}$$

( see[30]  p.305)

We define the g component vector indices  $\Delta^a, \Delta^b$  corresponding to the Riemann constant by the following equation:

$$\Delta = \Delta^a + \Omega \Delta^b \tag{4.31}$$

The 2g+2 characteristics $[U_j]$ serve as a basis for the construction of all $4^g$ possible half integer characteristics. There is a one-to-one correspondence between these characteristics and partitions of the set $\overline{\Gamma} = \{1,2,....2g+2\}$ of indices of the branching points.  Now we set $e_{2g+2} = \infty$ and this is taken as the base point of the Abel map.  We are interested in non-singular even characteristics.  We omit the point $\infty$ from consideration and are interested in the following partition of $\Gamma = \{1,2,....2g+1\}$ only:

$$\Gamma = A_0 \cup B_0 \qquad A_0 = \{i_1, i_2,...i_g\} \qquad B_0 = \{j_1, j_2,...j_{g+1}\} \tag{4.32}$$

There are  $\dfrac{1}{2}\dbinom{2g+2}{g+1} = \dbinom{2g+1}{g}$  different partitions, and the characteristic defined by

$$E \equiv \sum_{k=1}^{g} U^i_{j_k} + \Delta^i \tag{4.33}$$

for each of the partition is even.  We define the g component vector indices

$E^a, E^b$ by

$$E = E^a + \Omega E^b \tag{4.34}$$

The sigma function should be defined as [26][25], using the indices $\Delta^a, \Delta^b$ related to the Riemann constant ,

$$\sigma(u) = c \exp(-\frac{1}{2} \sum_{I,J=1}^{g} u_I \eta_{IJ} u_J) \theta \begin{bmatrix} \Delta^a \\ \Delta^b \end{bmatrix}(u) \tag{4.35}$$

Here the Jacobi theta function is defined in a standard notation:

$$\theta \begin{bmatrix} a \\ b \end{bmatrix}(z, \Omega) = \sum_{n \in Z^g} \exp\{i\pi(n+a)^t \Omega(n+a) + 2\pi i(n+a)(z+b)\} \tag{4.36}$$

By using a formula:

$$\theta \begin{bmatrix} a \\ b \end{bmatrix}(z, \Omega) = \theta(z + b + \Omega a, \Omega) \exp\{i\pi a^t \Omega a + 2\pi i a(z+b)\} \tag{4.37}$$

where $\theta(z, \Omega)$ is the standard theta function with zero index , we can show that

$$\partial_I \partial_J \ln \theta \begin{bmatrix} \Delta^a \\ \Delta^b \end{bmatrix}(z, \Omega) \Big|_{z=X^b + X^a \Omega} = \partial_I \partial_J \ln \theta \begin{bmatrix} \Delta^a + X^a \\ \Delta^b + X^b \end{bmatrix}(z, \Omega) \Big|_{z=0} \tag{4.38}$$

for any $X^a, X^b$. If we take $X^a, X^b$ as the index corresponding to $\sum_{k=1}^{g} U^i_{j_k}$, that is the index of the summation of half periods of chosen g branch points, the right hand side of (4.38) equals to, by (4.33),

$$\partial_I \partial_J \ln \theta \begin{bmatrix} E^a \\ E^b \end{bmatrix}(0, \Omega) = \frac{\partial_I \partial_J \theta \begin{bmatrix} E^a \\ E^b \end{bmatrix}(0)}{\theta \begin{bmatrix} E^a \\ E^b \end{bmatrix}(0)} \tag{4.39}$$

because $\partial_I \theta \begin{bmatrix} E^a \\ E^b \end{bmatrix}(0, \Omega) = 0$ for the even theta function. We denote this even index as $\delta$. The left hand side of is the value of $\partial_I \partial_J \ln \theta \begin{bmatrix} \Delta^a \\ \Delta^b \end{bmatrix}(z, \Omega)$ at a summation of half period corresponding to the chosen g branch points which is written as $\frac{1}{2}\Omega_\delta$.
Therefore, the meaning of the both sides of

$$P_{IJ}(\frac{1}{2}\Omega_\delta) \equiv \frac{1}{2}\eta_{IJ} - \partial_I\partial_J \ln \theta\begin{bmatrix} E^a \\ E^b \end{bmatrix}(0,\Omega) = \frac{1}{2}\eta_{IJ} - \frac{\partial_I\partial_J \theta[\delta](0)}{\theta[\delta](0)} \tag{4.40}$$

is now obvious.

In this way, if we can successfully decompose the product
$S_\delta(z_1,z_2)S_\delta(z_2,z_3)....S_\delta(z_M,z_1)$ into "constant part" and "function of the vertex inserting points part", we can attach a natural meaning to the former. We would also like to write the latter part totally in terms of sigma function, as was the case in g=1. That is, we would like to replace the odd theta function in (4.15) – (4.20) with the sigma function (4.35). This means that in the higher point amplitudes with M>=5 for g>1, the theta function of this form, $\partial_I\partial_J \ln \theta\begin{bmatrix} \Delta^a \\ \Delta^b \end{bmatrix}(z,\Omega)$ , or equivalently

$\partial_I\partial_J \ln \theta(z+\Delta,\Omega)$ ,will appear in the final form of the amplitudes after summing over spin structures of fermion field contractions. Something different from the Prime form, or something different form the bosonic Green functions will be one of the building components of the amplitudes.

Since the sigma function always appears in a form $\partial_I\partial_J \ln\sigma$ , the theta function of the form $\partial_I\partial_J \ln \theta\begin{bmatrix} \Delta^a \\ \Delta^b \end{bmatrix}(z,\Omega)$ can always be replaced with $\partial_I\partial_J \ln \theta(z+\Delta,\Omega)$ with the standard theta function, if we consider the formula (4.37) .

There occurs one problem: We can show that such theta function related to the Riemann constant is odd if $\frac{g(g+1)}{2}$ is odd, and even if $\frac{g(g+1)}{2}$ is even by simple calculations. Therefore, we have to restrict the whole of our considerations to the case that $\frac{g(g+1)}{2}$ is odd, since the theta function $\theta\begin{bmatrix} \Delta^a \\ \Delta^b \end{bmatrix}(z,\Omega)$ is originally from an odd theta function in the definition form of the Szego kernel. This is quite an ugly point ( or 'odd' point ) in our discussion. There may exist some reason that we can always use $\partial_I \partial_J \ln \theta(z+\Delta, \Omega)$ for the expression of Szego kernels, but this is an open issue.

One reason why we still think it may be appropriate to adopt $\partial_I \partial_J \ln \theta \begin{bmatrix} \Delta^a \\ \Delta^b \end{bmatrix}(z, \Omega)$ for any value of g in spite of this problem is as follows: g loop, M point amplitude will be modular invariant, and the final form of the fermion operator contraction terms after the summation over spin structure will be written by a unique, modular invariant function and constant modular forms. In genus1, this modular invariant function was $\frac{\partial^k}{\partial x^k} \ln \theta_1(x|\tau)$ and $\eta_1$, that is, the differentials of sigma function on torus. In the higher genus, this unique function will be the differentiation of sigma function defined above. The function form in the amplitude will be uniquely determined only by the pole structure when sigma function is used. This feature, which will make the modular invariance of M point g loop amplitude manifest, will be the case for the connected part too, and also even for non-hyper elliptic case too with appropriate definition of sigma function.

Anyway, at least for the case that $\frac{g(g+1)}{2}$ is odd, the Szego kernel may be written in a form analogous to the eq. (2.6)

$$S_\delta(z, w) = [\sum_{I,J=1}^{g} \{P_{IJ}(z-w) - P_{IJ}(\frac{1}{2}\Omega_\delta)\} \omega_I(z)\omega_J(w)]^{\frac{1}{2}} \quad (4.41)$$

Or, since Abel map is used in the notations,

$$S_\delta(z, w) = [-\partial_z \partial_w \ln \sigma(z-w) - \sum_{I,J=1}^{g} P_{IJ}(\frac{1}{2}\Omega_\delta) \omega_I(z)\omega_J(w)]^{\frac{1}{2}} \quad (4.42)$$

where, the use of the standard theta function which includes Riemann constant $\Delta$ in the form of Pe function is understood.

Eq. (4.17) is already a kind of non-trivial identity of decomposition which says that

$$S_\delta(z_1,z_2)S_\delta(z_2,z_3)S_\delta(z_3,z_1) = [\sum_{I,J=1}^{g} \{P_{IJ}(z_1-z_2) - P_{IJ}(\frac{1}{2}\Omega_\delta)\} \omega_I(z_1)\omega_J(z_2)]^{\frac{1}{2}}$$

$$\times [\sum_{K,L=1}^{g} \{P_{KL}(z_2-z_3) - P_{KL}(\frac{1}{2}\Omega_\delta)\} \omega_K(z_2)\omega_L(z_3)]^{\frac{1}{2}}$$

$$\times [\sum_{M,N=1}^{g} \{P_{MN}(z_3-z_1) - P_{MN}(\frac{1}{2}\Omega_\delta)\} \omega_M(z_3)\omega_N(z_1)]^{\frac{1}{2}} \}$$

$$= \text{ right hand side of (4.17)} \overset{schematically}{=} {}_3V_3^g + \sum_{comb\, I,J,K} {}_3V_1^g \cdot P_{IJ}(\frac{1}{2}\Omega_\delta) \quad (4.43)$$

In g=1, the square root disappeared as in (2.30) for any values of M.

Let us discuss how the constants $P_{IJ}(\frac{1}{2}\Omega_\delta)$ are determined at a fixed spin structure and how to sum over them. As is briefly described in the Appendix C, there exists a classical theorem corresponding to the determination of Pe function values at the summation of arbitrary g points on the hyper elliptic curve. See for example, chapter XI of [25] and [26].

Let $\quad u = \sum_{I=1}^{g} \int_{\infty}^{(x_I,y_I)} \omega \quad$, where $(x_I, y_I)$ are any g points on the hyper elliptic curve. Then, the following relationship holds:

$$\sum_{I=1}^{g}\sum_{J=1}^{g} P_{IJ}(u) x_r^{I-1} x_s^{J-1} = \frac{F(x_r, x_s) - 2 y_r y_s}{(x_r - x_s)^2} \quad (4.44)$$

for any choice of two points $x_r$, $x_s$, and

$$x_r^g - \sum_{J=1}^{g} P_{Jg}(u) x_r^{J-1} = 0 \quad (4.45)$$

for $r = 1,2,....g$.

Here, $F(x,z)$ is a well known polynomial, which is used in the realization of the Kleinian bi-differential $d\omega(x,y,z,w)$. Its explicit form is ( See Appendix C)

$$F(x,z) = \sum_{i=0}^{g} x^i z^i (\mu_{2g-2i}(x+z) + 2\mu_{2g-2i+1}) \quad (4.46)$$

$$= \{(x+z) + 2\mu_1\}x^g z^g + \{\mu_2(x+z) + 2\mu_3\}x^{g-1}z^{g-1} + ... + \mu_{2g}(x+z) + 2\mu_{2g+1} \quad (4.47)$$

where $\mu$ is defined in the definition of the curve :

$$y^2 = x^{2g+1} + \mu_1 x^{2g} + \mu_2 x^{2g-1} + .... + \mu_{2g+1} \quad (4.48)$$

With $\mu_1 = -\sum_{i=1}^{2g+1} e_i$ , $\mu_2 = +\sum_{i<j} e_i e_j$ , ..... (4.49)

Our interests are in each of the branch points $e_i$ corresponds to the point $(e_i, 0)$, and its half period is given by $\int_{\infty}^{(e_i,0)} \omega$ . Therefore, applying the theorem above for the chosen g number of branch points $x_1, x_2, ....x_g$ out of 2g+1 points, we immediately have

$$\sum_{I=1}^{g} \sum_{J=1}^{g} P_{IJ}(\frac{1}{2}\Omega_\delta) x_r^{I-1} x_s^{J-1} = \frac{F(x_r, x_s)}{(x_r - x_s)^2}$$ (4.50)

for any choice of two points $x_r$, $x_s$ ,and

$$x_r^g - \sum_{J=1}^{g} P_{Jg}(\frac{1}{2}\Omega_\delta) x_r^{J-1} = 0$$ (4.51)

for $r = 1, 2, ....g$.

From these two, we can show that if one of the index is equal to g,

" $(-1)^{g-I} P_{Ig}(\frac{1}{2}\Omega_\delta)$ is the degree $g-I+1$ fundamental symmetric function of $x_1, x_2, ....x_g$ ". (4.52)

What this theorem says is as follows. The $\frac{g(g+1)}{2}$ numbers of constants Pe function values at the summation of half periods of corresponding to the g points $x_1, x_2, ....x_g$ for a fixed spin structure, $P_{IJ}(\frac{1}{2}\Omega_\delta)$, can be determined in the following way. The g of them, $P_{Ig}$ $(I = 1, 2, ..g)$ are simply the fundamental symmetric function of $x_1, x_2, ....x_g$ . The rest $\frac{g(g-1)}{2}$ values are determined by a set of equations (4.50), each of which corresponds to choosing 2 branch points out of the chosen g points $x_1, x_2, ....x_g$ , whose number is $\binom{g}{2} = \frac{g(g-1)}{2}$ . This fact itself has nothing to do with string amplitudes, it is a result of the classical algebraic geometry. This theorem is naturally applied when one of the branch points is fixed at ∞ and we consider choosing g number of half periods out of 2g+1. Physically, this can be regarded as a general method to sum over spin structures in the disconnected parts of string amplitudes. String amplitudes in hyper elliptic case will have structures in which this simple and elegant method can be applied to sum up spin structures for arbitrary numbers of external bosons in any genus.

We also need to multiply the appropriate string measure. The searching of the correct measure, which has been one of the important topics in this area, was solved for the hyper elliptic case by[5]. They showed the explicit result as rational functions of 2g+2 ramification points. Here we need it in a slightly different form, using 2g+1 ramification points[4]. An example will be shown for the 2 loop case in the next section. After setting all of these, we do summation over spin structures in $P_{IJ}(\frac{1}{2}\Omega_\delta)$ expressed by $x_1, x_2, ....x_g$ ( and all of $e_1, e_2, .....e_{2g+1}$). This process will be done totally algebraically as we saw in g=1 case, just summing over all possibilities of $x_1, x_2, ....x_g$. As we saw, the cosmological constant, two point function, and three point function are zero because up to 3 product of Szego kernels contain only degree 0 and degree 1 polynomial of $P_{IJ}(\frac{1}{2}\Omega_\delta)$ .

In M point calculation, we will probably need a summation of the form $\sum_\delta measure(\delta) * Q(P_{IJ}(\frac{1}{2}\Omega_\delta))$ ,where $Q$ is degree $M-2$ polynomial of $P_{IJ}(\frac{1}{2}\Omega_\delta)$. After the summation, we will be left with manifestly modular invariant functions with appropriate poles of z1, z1, …zM, multiplied by symmetric functions of $e_1, e_2, .....e_{2g+1}$ .

These symmetric functions will correspond to genus g modular forms of theta constants. This part of consideration probably needs knowledge about the modular forms unknown at present.

---

[4] The auther does not know the exact proof of the general result in this slightly different formalism. Therefore its explicit form is not described here. In any case, it is essentially obtained in ref.[5].

# 5. Examples at genus 2

As an example at g=2, the curve is

$$y^2 = \prod_{k=1}^{5}(x-e_k) = R(x) = x^5 + \mu_1 x^4 + \mu_2 x^3 + \ldots + \mu_5 \qquad (5.1)$$

The function $F(x_1, x_2)$ is given by

$$F(x_1, x_2) = \{(x_1+x_2) + 2\mu_1\} x_1^2 x_2^2 + \{\mu_2(x_1+x_2) + 2\mu_3\} x_1 x_2 + \mu_4(x_1+x_2) + 2\mu_5 \quad (5.2)$$

Suppose that we adopt two points $x_1, x_2$ out of five points $e_1, e_2, \ldots e_5$.

For a fixed spin structure $\delta$, the equation (4.50) gives only one relationship:

$$P_{11} + P_{12}(x_1+x_2) + P_{22} x_1 x_2 = \frac{F(x_1, x_2)}{(x_1-x_2)^2} \qquad (5.3)$$

On the other hand, eq.(4.51) gives

$$x_1^2 - \{P_{12} + P_{22} x_1\} = 0 \quad , \qquad x_2^2 - \{P_{12} + P_{22} x_2\} = 0 \qquad (5.4)$$

Then we have the following solution, at a fixed spin structure,

$$P_{22} = x_1 + x_2 \quad , \quad P_{12} = P_{21} = -x_1 x_2 \quad , \quad P_{11} = \frac{F(x_1, x_2)}{(x_1-x_2)^2} \qquad (5.5)$$

It can be shown that $F(x_1, x_2)$ always contains a factor $(x_1-x_2)^2$ as in Appendix C, therefore all of $P_{11}, P_{12}, P_{22}$ are polynomials of $e_1, e_2, \ldots e_5$. Actually, in genus 2, if

$$x_1 = e_1, \ x_2 = e_2, \text{ then } \frac{F(x_1, x_2)}{(x_1-x_2)^2} = (e_3 + e_4 + e_5) x_1 x_2 + e_3 e_4 e_5 \ . \qquad (5.6)$$

Obviously the degrees of $P_{11}, P_{12}, P_{22}$ are of 3, 2, 1 respectively as functions of branch points.

Before going to exemplify some calculations of genus 2, let us describe something which can be said for arbitrary genus. The right hand side of general equation (4.50) is of degree $2g-1$. This is because F is of degree 2g+1 and F always contains the factor $(x_r - x_s)^2$. The degree of the $P_{IJ}$ are in general as follows:

$$\begin{pmatrix} P_{11} & P_{12} & P_{13} & \cdots & P_{1g} \\ P_{21} & P_{22} & P_{23} & \cdots & P_{2g} \\ \vdots & \vdots & \vdots & & \vdots \\ \vdots & \vdots & \vdots & & \vdots \\ P_{g1} & P_{g2} & P_{g3} & \cdots & P_{gg} \end{pmatrix} \overset{DEGREE}{\approx} \begin{pmatrix} 2g-1 & 2g-2 & 2g-3 & \cdots & g \\ 2g-2 & 2g-3 & 2g-4 & \cdots & g-1 \\ \vdots & \vdots & \vdots & & \vdots \\ \vdots & \vdots & \vdots & & \vdots \\ g & g-1 & g-2 & \cdots & 1 \end{pmatrix}$$

(5.7)

As is noted in Appendix D, the measure obtained in [5] has the form, in our normalization as in (1.5),

$$\frac{1}{\sqrt{D}}[\deg ree\ (2g-1)g\ polynomial\ of\ e_1.e_2,....e_{2g+1}] \tag{5.8}$$

where $\sqrt{D} \equiv \prod_{i<j}^{g} (e_i - e_j)$ is the square root of the discriminant of the curve. The

$\sqrt{D}$ is in general a polynomial of degree $\binom{2g+1}{2} = (2g+1)g$.

Any calculations are done by multiplying a function of $P_{IJ}$ to (5.8):

$$\sum \frac{1}{\sqrt{D}}[\deg ree\ (2g-1)g\ polynomial\ of\ e_1.e_2,....e_{2g+1}] \cdot (a\ function\ of\ P_{11}, P_{12}, P_{22})$$

(5.9)

and after summing over spin structures the numerator will contain the denominator $\sqrt{D}$ and the result will be symmetric function of $e_1.e_2,....e_{2g+1}$. For the M=0,2,3 point functions, the multiplied function to (5.8) is linear in $P_{IJ}$, whose highest degree is $2g-1$ of $P_{11}$. Then the argument given by A. Morozov in[29] can be applied, and will have vanishing result for M=0,2,3 point functions, because the degree of the numerator is not enough to give non-zero result. The function of $P_{11}, P_{12}, P_{22}$ multiplied in (5.9) should have degree 2g or higher to give non zero result. This was already seen in the case in g=1. Note that we have to do calculations only on the moduli parts of the amplitudes which have no dependence on the vertex inserting points z1, z2, ...zM, for any higher point calculations, and so the calculation will be easier.

It is very instructive to see the calculation of the part of the four point amplitude at genus 2, that is, a possible contraction from the $S_\delta(z_1,z_2)^2 S_\delta(z_3,z_4)^2$ .

Since $S_\delta(z_1,z_2)^2 = -\sum_{I,J=1}^{2} [\partial_I \partial_J \ln\theta(z_1-z_2-\Delta) - \frac{1}{2}\eta_{IJ} + P_{IJ}(\frac{1}{2}\Omega_\delta)] \omega_I(z)\omega_J(w)$

(5. 10)

only the following 4 terms x 4 terms product has the possibility to give non-zero result after summing over spin structures:

$$[\sum_{I,J=1}^{2} P_{IJ}(\delta)\, \omega_I(z_1)\omega_J(z_2)][\sum_{K,L=1}^{2} P_{KL}(\delta)\, \omega_K(z_3)\omega_L(z_4)] \quad (5.11)$$

The power counting of the degree of terms are the following:

$P_{22} \cdot P_{22}$      degree 2
$P_{22} \cdot P_{12}$      degree 3
$P_{12} \cdot P_{12}$      degree 4
$P_{11} \cdot P_{22}$      degree 4
$P_{12} \cdot P_{11}$      degree 5
$P_{11} \cdot P_{11}$      degree 6

To have non-zero result, the degree should be equal to or be more than 2g, therefore the last four have the possibility to give non-zero results.

As in the Appendix D of the DHP paper [1], what we have to calculate is:

$$\frac{(x_1-x_2)^4 \prod_{k\notin x_1,x_2}(e_k-x_1)(e_k-x_2) \prod_{k,l\notin x_1,x_2}(e_k-e_l)^2}{D} \sum_{i\neq 1,2} S_\delta(z_1,z_2)^2 S_\delta(z_3,z_4)^2 \quad (5.12)$$

and other combinations for the pair of x.   Here, k, l are from 1 to 5 .   We should be careful about the relative sign of each term of this measure part when $e_6$ is fixed to be $\infty$ , as described in Appendix D, eq.(D.5).   Note also one more summation in front of the products of Szego kernels, which excludes branch points $x_1, x_2$ when (5.11) is inserted.

The numerator is from famous $\Xi_6[\delta]\mathcal{9}[\delta]^4(\Omega)$ and the denominator comes from $\Psi_{10}(\Omega)$ in Ref. [1].   This is easily modified as

$$\frac{1}{\sqrt{D}}[(x_1-x_2)^3 \prod_{k,l\notin x_1,x_2}(e_k-e_l)\,]\sum_{i\neq 1,2} S_\delta(z_1,z_2)^2 S_\delta(z_3,z_4)^2 \quad (5.13)$$

The explicit numerical calculation shows the following results:

$$\sum_{spin} P_{12} \cdot P_{12} = 2\prod_{i<j}(e_i - e_j) = 2\sqrt{D} \tag{5.14}$$

$$\sum_{spin} P_{11} \cdot P_{22} = -4\sqrt{D} \tag{5.15}$$

$$\sum_{spin} P_{12} \cdot P_{11} = \sum_{spin} P_{11} \cdot P_{11} = 0 \tag{5.16}$$

Therefore, only $P_{12} \cdot P_{12}$ and $P_{11} \cdot P_{22}$ gives non-zero constant answer. This result, if substituted into eq. ( 5.11 ), shows the exact matching of the combinations of the biholomorphic 1 form [1]:

$$\Delta(x,u)\Delta(y,v) + \Delta(x,v)\Delta(y,u) \tag{5.17}$$

where $\quad \Delta(x,y) = \omega_1(x)\omega_2(y) - \omega_2(x)\omega_1(y) \tag{5.18}$

by our method of calculations, up to overall factor. When we calculate 5 point, 6 point amplitudes, the symmetric function of $e_1, e_2, \ldots e_5$ will remain non-zero, and will give modular forms in genus 2.

## 6. Discussions

The crucial point on which we have discussed in this document is the next observation:
The simple product of M Szego kernels $\prod_{i=1}^{M} S_\delta(z_i, z_{i+1})$ with the condition $z_{i+1} = z_1$ depends on the spin structures only through the one kind of constant $P_{IJ}(\frac{1}{2}\Omega_\delta)$ which is irrespective of $z_1, z_2, \ldots z_M$.

This is a fact for any M in g=1, and for M=1,2,3 for any g; and a conjecture for M>3 in general g. Under the situation this feature holds, we can use a result of classical theory of Abelian functions, which can be regarded as a general method of taking summation over spin structures systematically. This feature will hold even for non-hyper elliptic case.

There are some future problems remained unsolved.
The first is whether the decomposition feature as we saw in g=1 (2.18) holds or not in g>1 , M>3. If a generalized form of (2.18) is proved, the assumption of the amplitude dependence on the constant $P_{IJ}(\frac{1}{2}\Omega_\delta)$ is most beautifully confirmed for hyper elliptic cases. The genus 1 result (2.18) has a suggestive form、including the number of indices. The genus g result may have the form with K number of P

$$\prod_{i=1}^{M} S_\delta(z_i, z_{i+1}) \stackrel{?}{=} \sum_{K=0}^{\left[\frac{M}{2}\right]} \sum_{\text{comb } I_1, I_2, \ldots I_M} {}_M V_{M-2K}(z, I_1, I_2, \ldots I_{M-2K}) \cdot P_{I_{M-2K+1}, I_{M-2K+2}} \ldots P_{I_{M-1} I_M} \quad (6.1)$$

where $z_{M+1} = z_1$ and V are appropriate modular invariant functions.

The decomposition formula, if used with the method of summing over spin structures we described in section 4, principally will make it possible to calculate g loop, M point superstring amplitudes in an elegant way, though it is restricted to the case of the hyper elliptic disconnected parts. To prove such decomposition theorem, it may be not a good way to make use of Fay's formula only, and we may need a generalized Frobenius – Stickelberger type formula in higher genus.

The second is to seek whether we can say something on the connected parts of the amplitudes. There may be some more mathematical structures behind that parts too.

We saw a technical problem from the fact that the theta indices from the Riemann constant are sometimes even. On this point, probably there will be some natural remedy. The eq.(4.17) and (4.41), which are natural generalizations of those in g=1, will hold in any genus. An observation written above the (4.41) may be another reason of our expectation. The differentiation of sigma function in higher genus is a very good candidate for the unique function which shows good modular transformation properties.

Another direct application of the method described in this document would be a calculation of the disconnected parts of the g loop 4 point amplitude in hyper elliptic case.

From the contraction of the type $S_\delta(z_1, z_2)^2 S_\delta(z_3, z_4)^2$, it is obvious that we have non zero result only from the following part which is proportional to the 4 product of holomorphic 1 forms:

$$[\sum_{I,J=1}^{g} P_{IJ}(\delta) \, \omega_I(z_1) \omega_J(z_2)][\sum_{K,L=1}^{g} P_{KL}(\delta) \, \omega_K(z_3) \omega_L(z_4)] \quad (6.1)$$

From the contraction of the term $S_\delta(z_1, z_2) \, S_\delta(z_2, z_3) \, S_\delta(z_3, z_4) \, S_\delta(z_4, z_1)$, the method described in the Appendix D of ref.[1] may be applied.

By the power counting, there are many candidates of non-zero contributions in (6.1).

It would be very interesting to perform the summations over spin structure and see the explicit form of combinations $P_{IJ} \cdot P_{KL}$ the expansion of (6.1). If one can use good software, it may be possible to perform this calculation, since all are linear algebras. In any case, in the g loop 4 point amplitude in our restricted case, only the holomorphic 1 form will appear from the summation over spin structures of the fermion field contractions.

Even for general, non hyper elliptic string amplitudes, it will be useful to consider the structures of amplitudes totally in terms of the sigma function and its derivatives. The final form of fermion field contractions of the g loop ,M point amplitudes may have the following simple form: It will be constructed only by the differentials of log of genus g sigma function (4.35) and constant modular forms, both of which are manifestly modular invariant. If the inserting points of vertex operators are z1,z2, …zM, then the subsets of differences of those M points determines the differentials of log of genus g sigma function almost uniquely, as we exemplified in case of g=1.

# Appendix A  Genus 1  notations

The $\sigma$ function is defined as follows:

$$\sigma(z) = \sigma(z|2\omega_1, 2\omega_3) = z \prod_{m,n}{}' (1 - \frac{z}{\Omega_{m,n}}) \exp[\frac{z}{\Omega_{m,n}} + \frac{z^2}{2\Omega_{m,n}^2}] \tag{A.1}$$

Here,

$$\Omega_{m,n} = 2m\omega_1 + 2n\omega_3$$

In the following, we set $2\omega_1 = 1$, $2\omega_3 = \tau$ as is often denoted in the string theories. Some standard, classical functions are defined as

$$\varsigma(z) = \frac{d \ln \sigma(z)}{dz} \qquad P(z) = -\frac{d\varsigma(z)}{dz}$$

$$\omega_2 \equiv -(\omega_1 + \omega_3) \qquad e_\delta = P(\omega_\delta) \qquad \eta_\delta = \varsigma(\omega_\delta) \quad (\delta = 1,2,3)$$

$$g_2 = 60 \sum_{m,n} \Omega_{m,n}^{-4} \qquad g_3 = 140 \sum_{m,n} \Omega_{m,n}^{-6} \qquad S_\delta(z) = \{P(z) - e_\delta\}^{1/2} \tag{A.2}$$

$$P^{(2n)}(z) = \frac{d^{2n} P(z)}{dz^{2n}} = \text{polynomial of } P(z) \text{ of order } n+1 \tag{A.3}$$

$$P^{(2n+1)}(z) = P'(z) * [\text{polynomial of } P(z) \text{ of order } n] \tag{A.4}$$

$$P'(\omega_\delta) = 0 \tag{A.5}$$

$$\{P'(z)\}^2 = 4\{P(z)\}^3 - g_2 P(z) - g_3 \tag{A.6}$$

$$\sigma(2\omega_1 z) = 2\omega_1 \exp(2\eta_1 \omega_1 z^2) \frac{\theta_1(z)}{\dot{\theta}_1(0)} \qquad \eta_1 = \varsigma(\omega_1) = -\frac{1}{6}\frac{\dddot{\theta}_1(0)}{\dot{\theta}_1(0)} \tag{A.7}$$

$$P_\delta(2\omega_1 z) = \frac{1}{2\omega_1} \frac{\dot{\theta}_1(0) \theta_{\delta+1}(z)}{\theta_{\delta+1}(0) \theta_1(z)} \tag{A.8}$$

$$e_1 + e_2 + e_3 = 0 \qquad e_1 e_2 + e_2 e_3 + e_3 e_1 = -\frac{g_2}{4} \qquad e_1 e_2 e_3 = \frac{g_3}{4} \tag{A.9}$$

The $g_2, g_3$ is obviously proportional to Eisenstein series $E_4, E_6$ as in (A.2).

$$e_2 - e_1 = -\frac{\pi^2 \theta_4^4(0)}{4\omega_1^2} \quad e_3 - e_2 = -\frac{\pi^2 \theta_2^4(0)}{4\omega_1^2} \quad e_1 - e_3 = \frac{\pi^2 \theta_3^4(0)}{4\omega_1^2} \tag{A.10}$$

That is,

$$2\omega_1 = 1 = \frac{\pi \theta_4^2(0)}{\sqrt{e_1 - e_2}} = \frac{\pi \theta_3^2(0)}{\sqrt{e_1 - e_3}} = \frac{\pi \theta_2^2(0)}{\sqrt{e_2 - e_3}} \tag{A.11}$$

This is an example of Thomae type formulas. From this, together with the Picard-Fuchs equations, Legendre's relations, and the heat equation, we can derive[24]

$$\eta_1 = \varsigma(\omega_1) = -\frac{1}{12\omega_1}\frac{\dddot{\theta}_1(0)}{\dot{\theta}_1(0)} = -\omega_1 e_1 - \frac{1}{4\omega_1}\frac{\ddot{\theta}_2(0)}{\theta_2(0)} = -\omega_1 e_2 - \frac{1}{4\omega_1}\frac{\ddot{\theta}_3(0)}{\theta_3(0)} = -\omega_1 e_3 - \frac{1}{4\omega_1}\frac{\ddot{\theta}_4(0)}{\theta_4(0)} \tag{A.12}$$

Therefore,

$$e_\delta = \frac{\ddot{\theta}_{\delta+1}(0)}{\theta_{\delta+1}(0)} - 2\eta_1 = \frac{\ddot{\theta}_{\delta+1}(0)}{\theta_{\delta+1}(0)} - \frac{1}{3}\frac{\dddot{\theta}_1(0)}{\dot{\theta}_1(0)} \quad (\delta = 1,2,3) \tag{A.13}$$

and consequently,

$$\frac{\ddot{\theta}_3(0)}{\theta_3(0)} - \frac{\ddot{\theta}_2(0)}{\theta_2(0)} = \pi^2 \theta_4^4(0), \quad \frac{\ddot{\theta}_4(0)}{\theta_4(0)} - \frac{\ddot{\theta}_3(0)}{\theta_3(0)} = \pi^2 \theta_2^4(0), \quad \frac{\ddot{\theta}_2(0)}{\theta_2(0)} - \frac{\ddot{\theta}_4(0)}{\theta_4(0)} = \pi^2 \theta_3^4(0). \tag{A.14}$$

Other famous formulas are

$$\frac{\dddot{\theta}_1(0)}{\dot{\theta}_1(0)} - \left[\frac{\ddot{\theta}_2(0)}{\theta_2(0)} + \frac{\ddot{\theta}_3(0)}{\theta_3(0)} + \frac{\ddot{\theta}_4(0)}{\theta_4(0)}\right] = e_1 + e_2 + e_3 = 0 \tag{A.15}$$

$$\dot{\theta}_1(0) = \pi \theta_2(0) \theta_3(0) \theta_4(0) \tag{A.16}$$

Dedekind function

$$\eta^{24} = \pi^{12}\{\theta_2(0)\theta_3(0)\theta_4(0)\}^8 = (e_1 - e_2)^2(e_2 - e_3)^2(e_3 - e_1)^2 \tag{A.17}$$

In the following, we prove that
$$S_\delta(x_1)S_\delta(x_2)S_\delta(x_3) = {}_3V_3(x_1,x_2,x_3) + {}_3V_1(x_1,x_2,x_3)e_\delta \qquad (2.17)$$
is equivalent to the Fay's trisecant identity. We derive the Fay's formula starting from (2.17) which can be derived without Fay's formula as in the proof of theorem in the section 2.

Consider to calculate
$$S_\delta(z_1 - w_2)S_\delta(z_2 - w_1)S_\delta(z_1 + z_2 - w_1 - w_2) - S_\delta(z_1 - w_2)S_\delta(z_2 - w_1)S_\delta(z_1 + z_2 - w_1 - w_2)$$

If we consider eq.(2.17), this is modified to the form of $G + He_\delta$, and obviously the zeros of both of G and H are at $z_1 = z_2$ $w_1 = w_2$, and H has another zero at $z_1 + z_2 = w_1 + w_2$. The poles are easily seen from the subtractions. Then, we can write,

$$G = \frac{E(z_1 - z_2)E(w_1 - w_2)P(z_1 + z_2 - w_1 - w_2)E(z_1 + z_2 - w_1 - w_2)}{\dot\theta_1(0)E(z_1 - w_1)E(z_2 - w_1)E(z_1 - w_2)E(z_2 - w_2)} \qquad (A.18)$$

$$H = -\frac{E(z_1 - z_2)E(w_1 - w_2)E(z_1 + z_2 - w_1 - w_2)}{\dot\theta_1(0)E(z_1 - w_1)E(z_2 - w_1)E(z_1 - w_2)E(z_2 - w_2)} \qquad (A.19)$$

where, in G, we made the pole of the first order by combining
$$P(z_1 + z_2 - w_1 - w_2)E(z_1 + z_2 - w_1 - w_2)$$

Using $P(z_1 + z_2 - w_1 - w_2) - e_\delta = S_\delta(z_1 + z_2 - w_1 - w_2)^2$ we have

$$S_\delta(z_1 - w_2)S_\delta(z_2 - w_1)S_\delta(z_1 + z_2 - w_1 - w_2) - S_\delta(z_1 - w_2)S_\delta(z_2 - w_1)S_\delta(z_1 + z_2 - w_1 - w_2)$$
$$= \frac{E(z_1 - z_2)E(w_1 - w_2)S_\delta(z_1 + z_2 - w_1 - w_2)^2 E(z_1 + z_2 - w_1 - w_2)}{\dot\theta_1(0)E(z_1 - w_1)E(z_2 - w_1)E(z_1 - w_2)E(z_2 - w_2)}$$

This is equivalent to the Fay's formula.

The theorem of section 2 is trivially re-written as

$$S_\delta(x_1 + x_2 + \ldots + x_{M-1}) = \frac{1}{\prod_{i=1}^{M-1} S_\delta(x_i)} \cdot \sum_{K=0}^{[\frac{M}{2}]} \left[{}_M V_{M-2K}(x_1, x_2, \ldots x_M)\right] \cdot (e_\delta)^K \qquad (\delta = 1, 2, 3)$$

We call this an addition theorem of Szego kernel at g=1. In this form, $x_M$ is defined by $\sum_{i=1}^{M} x_i = 0$. In the above it is proved that for M=3 this identity is equivalent to Fay's formula, but it is not obvious to show the same equivalence for M>3. It seems difficult to derive the theorem directly from Fay's identities.   This identity may contain new information for M>3.

## Appendix B   Some formulas of determinants

### B-1   Fundamental notations and formulae

In the following,   $y_1, y_2, \cdots y_n$ are   n   complex numbers.
First, define the determinant   $G(k, y_1, y_2, \cdots y_n)$   as follows:

$$G(k, y_1, y_2, \cdots y_n) \equiv \begin{vmatrix} 1 & y_1 & y_1^2 & \cdots & y_1^{k-1} & y_1^{k+1} & \cdots & y_1^n \\ 1 & y_2 & y_2^2 & \cdots & y_2^{k-1} & y_2^{k+1} & \cdots & y_2^n \\ \vdots & \vdots & \vdots & & \vdots & \vdots & & \vdots \\ \vdots & \vdots & \vdots & & \vdots & \vdots & & \vdots \\ 1 & y_n & y_n^2 & \cdots & y_n^{k-1} & y_n^{k+1} & \cdots & y_n^n \end{vmatrix} \tag{B.1}$$

It slightly differs from Van der Monde type; the k th degree of the variable is absent. The case of k=1 will be used later:

$$G(1, y_1, y_2, \cdots y_n) \equiv \begin{vmatrix} 1 & y_1^2 & y_1^3 & \cdots & & \cdots & y_1^n \\ 1 & y_2^2 & y_2^3 & \cdots & & \cdots & y_2^n \\ \vdots & \vdots & \vdots & & & & \vdots \\ \vdots & \vdots & \vdots & & & & \vdots \\ 1 & y_n^2 & y_n^3 & \cdots & & \cdots & y_n^n \end{vmatrix} \tag{B.2}$$

This form of   determinant appears in the denominator of the ratio   $\dfrac{a_j}{a_{M-2}}$.   This corresponds to the fact that Pe function is a second order elliptic function, and so there is no one order variable in the inverse of the poles.

One more thing we need to define is the following   $L(1, k, y_1, y_2, \cdots y_n)$   :

$$L(1, k, y_1, y_2, \cdots y_n) \equiv \begin{vmatrix} 1 & y_1^2 & y_1^3 & \cdots & y_1^{k-1} & y_1^{k+1} & \cdots & y_1^{n+1} \\ 1 & y_2^2 & y_2^3 & \cdots & y_2^{k-1} & y_2^{k+1} & \cdots & y_2^{n+1} \\ \vdots & \vdots & \vdots & & \vdots & \vdots & & \vdots \\ \vdots & \vdots & \vdots & & \vdots & \vdots & & \vdots \\ 1 & y_n^2 & y_n^3 & \cdots & y_n^{k-1} & y_n^{k+1} & \cdots & y_n^{n+1} \end{vmatrix} \tag{B.3}$$

Also, we define $_nW_k(y_1, y_2, \cdots y_n)$ as follows :

$_nW_k(y_1, y_2, \cdots y_n) \equiv$ A polynomial constructed as follows: Make a product of k numbers of variables out of n number of variables, and add those products symmetrically over all possible ways. We define W=1 when k=0.

For example,
$$_3W_1(y_1, y_2, y_3) \equiv y_1 + y_2 + y_3$$
$$_3W_2(y_1, y_2, y_3) \equiv y_1 y_2 + y_2 y_3 + y_3 y_1$$
$$_3W_3(y_1, y_2, y_3) \equiv y_1 y_2 y_3$$

$_nW_{n-1}$ consists of n terms, and is the sum of the products of (n-1) numbers out of n.

$_nW_n$ consists of only one term, which is all products of n variables.

Then, by elementary calculations, we can prove that :

Formula 1 : $G(k, y_1, y_2, \cdots y_n) = {}_nW_{n-k}(y_1, y_2, \cdots y_n) \cdot \prod_{i<j}(y_i - y_j)$ (B.4)

That is, G is the product of Van der Monde determinant and W.

Further,

Formula 2 : $L(1, k, y_1, y_2, \cdots y_n) = ({}_nW_{n-1} \cdot {}_nW_{n+1-k} - {}_nW_n \cdot {}_nW_{n-k}) \cdot \prod_{i<j}(y_i - y_j)$ (B.5)

The simplest example of the formula 2 is

$$L(1,3, y_1, y_2, y_3) = \begin{vmatrix} 1 & y_1^2 & y_1^4 \\ 1 & y_2^2 & y_2^4 \\ 1 & y_3^2 & y_3^4 \end{vmatrix} = (y_1 + y_2)(y_2 + y_3)(y_3 + y_1) \cdot \prod_{i<j}(y_i - y_j)$$

$$= \{(y_1 + y_2 + y_3)(y_1 y_2 + y_2 y_3 + y_3 y_1) - y_1 y_2 y_3\} \cdot \prod_{i<j}(y_i - y_j) = ({}_3W_1 \cdot {}_3W_2 - {}_3W_3 \cdot {}_3W_0) \cdot \prod_{i<j}(y_i - y_j)$$

Here $_3W_0 = 1$ as defined above.

## B-2  Calculation of the ratio of determinants in $G_M$

We concentrate on the pole structure of this ratio of determinants. The simplest case is M=3:

$$\frac{a_0}{a_1} = \frac{\begin{vmatrix} 1 & P'(x_1) \\ 1 & P'(x_2) \end{vmatrix}}{\begin{vmatrix} 1 & P(x_1) \\ 1 & P(x_2) \end{vmatrix}} \cong \frac{\frac{1}{x_2^3} - \frac{1}{x_1^3}}{\frac{1}{x_2^2} - \frac{1}{x_1^2}} = \frac{1}{x_1} + \frac{1}{x_2} - \frac{1}{x_1 + x_2} = \frac{1}{x_1} + \frac{1}{x_2} + \frac{1}{x_3}$$, This is equal to

$$\frac{a_0}{a_1} = \frac{L(1,2,\frac{1}{x_1},\frac{1}{x_2})}{G(1,\frac{1}{x_1},\frac{1}{x_2})} = \frac{{}_2W_1 \cdot {}_2W_1 - {}_2W_2 \cdot {}_2W_0}{{}_2W_1}$$

This ratio of the determinants of Pe function is equal to $\sum_{i=1}^{3} \frac{\partial}{\partial x_i} \ln \theta_1(x_i)$ .

Let us calculate the ratio $\dfrac{a_2}{a_{M-2}}$ as the second example.

$M = 5$

$$\frac{a_2}{a_{M-2}} \cong \frac{L(1,4,\frac{1}{x_1},\frac{1}{x_2},\frac{1}{x_3},\frac{1}{x_4})}{G(1,\frac{1}{x_1},\frac{1}{x_2},\frac{1}{x_3},\frac{1}{x_4})} = \frac{{}_4W_3 \cdot {}_4W_1 - {}_4W_4 \cdot {}_4W_0}{{}_4W_3} = \sum_{i=1}^{4} \frac{1}{x_i} - \frac{1}{\sum_{i=1}^{4} x_i} \quad (B.6)$$

Considering $x_5 = -\sum_{i=1}^{4} x_i$ , the final result, the rightest hand of the above equation, is simply equal to $\sum_{i=1}^{5} \frac{1}{x_i}$ . That is, $\dfrac{a_2}{a_{M-2}}$ is the sum of elliptic functions with poles $\dfrac{1}{x_i}$ . At first sight the order of poles are one, but by the condition $\sum_{i=1}^{5} x_i = 0$ each variables are not independent, and the final result is elliptic.

$M = 6$

$$\frac{a_2}{a_{M-2}} \cong \frac{L(1,4,\frac{1}{x_1},\frac{1}{x_2},\frac{1}{x_3},\frac{1}{x_4},\frac{1}{x_5})}{G(1,\frac{1}{x_1},\frac{1}{x_2},\frac{1}{x_3},\frac{1}{x_4},\frac{1}{x_5})} = \frac{{}_5W_4 \cdot {}_5W_2 - {}_5W_5 \cdot {}_5W_1}{{}_5W_4}$$

$$= \sum_{i,j=1}^{5}{}' \frac{1}{x_i \cdot x_j} - (\sum_{i=1}^{5} \frac{1}{x_i}) \cdot \frac{1}{\sum_{i=1}^{5} x_i} = \sum_{i,j=1}^{6}{}' \frac{1}{x_i \cdot x_j} \quad (B.7)$$

Where the prime means that we exclude the combination of $i = j$.

$M = 7$

$$\frac{a_2}{a_{M-2}} \cong \frac{L(1, 4, \frac{1}{x_1}, \frac{1}{x_2}, \frac{1}{x_3}, \frac{1}{x_4}, \frac{1}{x_5}, \frac{1}{x_6})}{G(1, \frac{1}{x_1}, \frac{1}{x_2}, \frac{1}{x_3}, \frac{1}{x_4}, \frac{1}{x_5}, \frac{1}{x_6})} = \frac{{}_6W_5 \cdot {}_6W_3 - {}_6W_6 \cdot {}_6W_2}{{}_6W_5}$$

$$= \sum_{i,j,k=1}^{6}{}' \frac{1}{x_i \cdot x_j \cdot x_k} - (\sum_{i,j=1}^{6} \frac{1}{x_i \cdot x_j}) \cdot \frac{1}{\sum_{i=1}^{6} x_i} = \sum_{i,j,k=1}^{7}{}' \frac{1}{x_i \cdot x_j \cdot x_k} \qquad (B.8)$$

In general, for arbitrary value of M, the ratio of two determinants $\frac{a_j}{a_{M-2}}$ $(0 \leq j \leq M-2)$ has the following pole structures using the variables $x_i$ $(1 \leq i \leq M)$. This pole structure completely determines the form of this ratio as an elliptic function up to the over all constants:

$$\frac{a_j}{a_{M-2}} \approx \sum_{sums\,over\,all\,possibilies} \frac{1}{\text{Product of } M - j - 2 \text{ variables chosen out of } M \text{ number of } x_i}$$

Generally, ${}_MV_{M-2K}$ has the pole structure in terms of y or x:

$${}_MV_{M-2K} \approx \frac{L(1, 2k, y_1, y_2, ... y_{M-1})}{G(1, y_1, y_2, ... y_{M-1})} = \frac{{}_{M-1}W_{M-2} \cdot {}_{M-1}W_{M-2K} - {}_{M-1}W_{M-1} \cdot {}_{M-1}W_{M-1-2K}}{{}_{M-1}W_{M-2}}$$

$$= {}_{M-1}W_{M-2K} - \frac{{}_{M-1}W_{M-1} \cdot {}_{M-1}W_{M-1-2K}}{{}_{M-1}W_{M-2}} .$$

The ratio $\frac{{}_{M-1}W_{M-1}}{{}_{M-1}W_{M-2}}$ is $\frac{1}{\sum_{i=1}^{M-1} x_i}$, that is, $-\frac{1}{x_M}$. The meaning of ${}_{M-1}W_{M-2K}$ is the sum of M-2K number of products excluding $x_M$, and adding $\frac{1}{x_M} \cdot {}_{M-1}W_{M-1-2K}$ we have the whole of ${}_MW_{M-2K}$.

## Appendix C  Formulas on Kleinian bi-differential and genus g Pe functions

Let $A_1, A_2, ... A_g$ and $B_1, B_2, ... B_g$ be a canonical homology basis. We choose canonical holomorphic differentials of the first kind $\omega_1, \omega_2, ... \omega_g$ and associated meromorphic differentials of the second kind $r_1, r_2, ... r_g$. The periods are given as

$$\oint_{A_I} \omega_J = \delta_{IJ} \qquad (C.1)$$

$$\oint_{B_I} \omega_J = \Omega_{IJ} \qquad (C.2)$$

$$\oint_{A_I} r_J = \eta_{IJ} \qquad (C.3)$$

$$\oint_{B_I} r_J = \eta'_{IJ} \qquad (C.4)$$

Let $V(x, y)$ be the hyperelliptic curve given by the equation:

$$y^2 = \prod_{k=1}^{2g+1}(x - e_k) = R(x) = x^{2g+1} + \mu_1 x^{2g} + \mu_2 x^{2g-1} + \ldots + \mu_{2g+1} \qquad (C.5)$$

where a variable y instead of s of eq.(1.3) is used, and one of the branch points $e_{2g+2}$ is fixed at $\infty$. Needless to say, all of $\mu_1, \mu_2, \ldots \mu_{2g+1}$ are fundamental symmetric polynomial of $e_1, e_2, \ldots e_{2g+1}$.

A well known polynomial $F(x, z)$ is defined as follows:

$$F(x, z) = \sum_{i=0}^{g} x^i z^i (\mu_{2g-2i}(x + z) + 2\mu_{2g-2i+1})$$

$$= \{(x + z) + 2\mu_1\} x^g z^g + \{\mu_2(x + z) + 2\mu_3\} x^{g-1} z^{g-1} + \ldots + \mu_{2g}(x + z) + 2\mu_{2g+1} \qquad (C.6)$$

The Kleinian bi-differential $d\omega(x, y, z, w)$ on $V \times V$ can be realized as, using this polynomial $F(x, z)$,

$$d\omega(x, y, z, w) = \frac{2yw + F(x, z)}{4(x - z)^2} \frac{dx}{y} \frac{dz}{w} \qquad (C.7)$$

The following theorems are known, proved by the properties of the Kleinian bi-differential [25]. It is most beautifully proved by using sigma function in genus g [26].

<Theorem 1> Let $u = \sum_{I=1}^{g} \int_{\infty}^{(x_I, y_I)} \omega$ for any g number of points $(x_1, y_1), \ldots (x_g, y_g)$, on the curve.

Then the following relations hold, for any r:

$$2y_r = P_{ggg}x_r^{g-1} + P_{gg,g-1}x_r^{g-2} + \ldots + P_{gg2}x_r + P_{gg1} \tag{C.8}$$

$$x_r^g - \sum_{J=1}^{g} P_{Jg}(u)x_r^{J-1} = 0 \tag{C.9}$$

$$\sum_{I=1}^{g}\sum_{J=1}^{g} P_{IJ}(u)x_r^{I-1}x_s^{J-1} = \frac{F(x_r,x_s) - 2y_r y_s}{(x_r - x_s)^2} \tag{C.10}$$

for any choice of two points $(x_r, y_r), (x_s, y_s)$ ,

Here we defined $\quad P_{IJK} \equiv \partial_K P_{IJ}$ .

From these, it can be derived that

$(-1)^{g-J} P_{Jg}(u) =$ fundamental symmetric function of $x_1, x_2, \ldots x_g$ of order $g - J + 1$ .

The polynomial $F(x,z)$ has the following properties

$$F(x,x) = 2R(x) \quad , \quad \left.\frac{\partial}{\partial x}F(x,z)\right|_{x=z} = \frac{d}{dz}R(z) \tag{C.11}$$

where $R(x)$ is the curve itself defined in eq.(C.5)

If we expand the $F(x,z)$ around $x = z$ , we have

$$F(x,z) = F(z,z) + (x-z)\frac{d}{dz}R(z) + (x-z)^2 H(x,z) \tag{C.12}$$

where everything else is included in $H(x,z)$ . If $x_1, x_2$ are any of two branch points, then it is apparent from (C.5) and (C.11) that F has the form

$$F(x_1, x_2) = +(x_1 - x_2)\left.\frac{d}{dz}R(z)\right|_{z=x_2} + (x_1 - x_2)^2 H(x_1, x_2) \quad . \quad \text{and the factor } \left.\frac{d}{dz}R(z)\right|_{z=x_2}$$

contains one more factor $(x_1 - x_2)$. Therefore, the right hand side of eq.(4.50) is always polynomial.

The polynomial $\dfrac{F(x_1, x_2)}{(x_1 - x_2)^2}$ is obtained without much difficulty. For example, in g=2, let us we adopt $x \equiv x_1 = e_1$ and $x_2 = e_2$ . Assuming the form of $F(x, e_2) = (x - e_2)^2(Ax + B)$ , differentiate two times both of this equation and the (5.2). Then the coefficients A and B are easily determined. The result is

$$\frac{F(e_1, e_2)}{(e_1 - e_2)^2} = (e_3 + e_4 + e_5)e_1 e_2 + e_3 e_4 e_5 \quad .$$

# Appendix D  Miscellaneous Issues

The hyper elliptic measure obtained in [5] has the form, at a fixed spin structure,

$$H_g = \frac{1}{2^g} \frac{1}{g+1} (\Delta_{\{1,2,\ldots,2g+2\}})^2 \sum_{k \in \varepsilon_{2g+2}} \frac{-1}{\psi_k} \quad (D.1)$$

in Definition 1 or in Theorem 14 in [5].

Just for convenience, we repeat the notations written in [5] here.

Definition:

$$\Delta_T \equiv \prod_{i,j \in T; i > j} (e_i - e_j) \quad (D.2)$$

For a finite sequence of natural numbers $k = (k_1, \ldots, k_r)$ define $\psi'_k = \prod_{i=1}^{r-1} (e_{k_i} - e_{k_{i+1}})$

and $\psi_k = (e_{k_r} - e_{k_1}) \psi'_k$.

Define $\varepsilon_r = \{(k_1, \ldots, k_r) \in N^r : \forall i, k_i \equiv i \mod 2 \text{ and } \{k_1, \ldots, k_r\} = \{1, \ldots, r\}\}$  That is, $\varepsilon_r$ consists of all permutations of $(1, \ldots, r)$ that alternate odd and even, beginning with odd.  Then the $H_g$ is defined as in (D.1)

When $e_{2g+2}$ is fixed at $\infty$, the factors of pairs which contain $e_{2g+2}$ in (D.1) are replaced as $(e_i - e_{2g+2}) \to -1$, that is, pick up the sign in front of $e_{2g+2}$ and erase the factor.  This is known as star map (page3, [5]).

After this procedure is done, the $H_g$ is $H_g = \frac{1}{2^g} \frac{1}{g+1} D \sum_{k \in \varepsilon_{2g+2}} \frac{-1}{\psi_k}$, where we assumed star map is taken in $\psi_k$, and D is the discriminant of the curve (C.5).

In our notations, the string measure is $\frac{H_g}{D}$, and we do a trivial modification:

$$\frac{H_g}{D} = \frac{1}{2^g} \frac{1}{g+1} \frac{1}{\sqrt{D}} (-1) \sum_{k \in \varepsilon_{2g+2}} \frac{\sqrt{D}}{\psi_k} \quad (D.3)$$

The degree of $\psi_k$ is 2g, after two factors disappeared by the star map.  The degree of $\sqrt{D}$ is (2g+1)g, and therefore

$$\frac{H_g}{D} \approx \frac{1}{\sqrt{D}} [\deg ree\ (2g-1)g\ polynomial\ of\ e_1, e_2, \ldots e_{2g+1}] \tag{D.4}$$

When we use (5.12) or (5.13) in the actual calculations, we have to be careful about the relative signs among the measure factors coming from the star map explained above. In g=2, if we place the two pairs of e as $(e_i - e_j)$ for $i < j$, we have the following signs before each of the numerators of the measure parts (four + and six −) in (5.12), (5.13):

$$-(e_1 - e_2)^3 (e_3 - e_4)(e_3 - e_5)(e_4 - e_5)$$

$$+(e_1 - e_3)^3 (e_2 - e_4)(e_2 - e_5)(e_4 - e_5)$$

$$-(e_1 - e_4)^3 (e_2 - e_3)(e_2 - e_5)(e_3 - e_5)$$

$$+(e_1 - e_5)^3 (e_2 - e_3)(e_2 - e_4)(e_3 - e_4)$$

$$-(e_2 - e_3)^3 (e_1 - e_4)(e_1 - e_5)(e_4 - e_5)$$

$$+(e_2 - e_4)^3 (e_1 - e_3)(e_1 - e_5)(e_3 - e_5)$$

$$-(e_2 - e_5)^3 (e_1 - e_3)(e_1 - e_4)(e_3 - e_4)$$

$$-(e_3 - e_4)^3 (e_1 - e_2)(e_1 - e_5)(e_2 - e_5)$$

$$+(e_3 - e_5)^3 (e_1 - e_2)(e_1 - e_4)(e_2 - e_4)$$

$$-(e_4 - e_5)^3 (e_1 - e_2)(e_1 - e_3)(e_2 - e_3)$$

.(D.5)

For each of above terms, $\frac{1}{\sqrt{D}} \sum_{i \neq 1,2} S_\delta(z_1, z_2)^2 S_\delta(z_3, z_4)^2$ is multiplied and summed over.

A.G. Tsuchiya   E-mail Address    colasumi@theia.ocn.ne.jp